\documentclass[journal,10pt, twocolumn, final]{IEEEtran}
\pdfoutput=1

\usepackage[]{graphicx}    
\newcommand{\ignore}[1]{}  
\usepackage[dvipsnames, table]{xcolor}
\usepackage{hhline}
\usepackage[labelfont=bf, justification=centering]{caption}
\usepackage{subfigure}
\usepackage{float} 
\usepackage{multirow}
\usepackage{float}
\usepackage{url}
\graphicspath{{./figures/}}
\usepackage{amsmath}
\usepackage{soul}
\usepackage{comment}
\usepackage{balance}
\usepackage{gensymb}
\usepackage{cite}
\newcolumntype{M}[1]{>{\centering\arraybackslash}m{#1}}
 \usepackage[compact]{titlesec}
\titlespacing{\section}{0pt}{1ex}{0.5ex}
\titlespacing{\subsection}{0pt}{1ex}{0ex}
\titlespacing{\subsubsection}{0pt}{0.5ex}{0ex}
\usepackage{paralist}
\usepackage[bottom]{footmisc}
\usepackage{wrapfig}

\begin{document}

\title{Modeling Interference from Millimeter Wave and Terahertz Bands Cross-links in Low Earth Orbit Satellite Networks for 6G and Beyond}

\author{Sergi Aliaga, Vitaly Petrov, and Josep M. Jornet
\vspace{-9mm}
\thanks{The authors are with the Department of Electrical and Computer Engineering, Northeastern University, Boston, MA, USA. Email: \{aliaga.s, v.petrov, j.jornet\}@northeastern.edu.}
\thanks{A shorter version of this work has been presented at the IEEE ICC'23 in Rome, Italy in June~2023, available at \protect\url{https://arxiv.org/abs/2302.04169}~\cite{sergi_ICC_2023}.}}

\maketitle

\begin{abstract}

High-rate satellite communications among hundreds and even thousands of satellites deployed at low-Earth orbits (LEO) will be an important element of the forthcoming sixth-generation (6G) of wireless systems beyond 2030. With millimeter wave communications (mmWave, $\approx$$30$\,GHz--$100$\,GHz) completely integrated into 5G terrestrial networks, exploration of its potential, along with sub-terahertz (sub-THz, $100$\,GHz--$300$\,GHz), and even THz ($300$\,GHz--$3$\,THz) frequencies, is underway for space-based networks. However, the interference problem between LEO mmWave/THz satellite cross-links in the same or different constellations is undeservedly forgotten. This article presents a comprehensive mathematical framework for modeling directional interference in all key possible scenario geometries. The framework description is followed by an in-depth numerical study on the impact of cross-link interference on various performance indicators, where the delivered analytical results are cross-verified via computer simulations. The study reveals that, while highly directional mmWave and, especially, THz beams minimize interference in many cases, there are numerous practical configurations where the impact of cross-link interference cannot be neglected and must be accounted for.
\end{abstract}
 
\vspace{-3mm}
\section{Introduction}
\label{sec:intro}
One of the key advancements on the path from 5G through 5G-Advanced to 6G systems is the seamless integration of high-rate satellite communications into the cellular networking environment~\cite{Giordani2021non, Gavras2022Towards}. Specifically, massive constellations of low-Earth orbit (LEO) satellites are seen as crucial facilitators for pervasive, high-speed communication around the world~\cite{Lin2021On}. These near-Earth systems complete the ``connectivity triad" by complementing wireless local area networks and cellular networks in providing high-rate and low-latency Internet access for users on land, sea, and air~\cite{Zhu2022IoT}.

To reach the anticipated 6G-grade performance levels, the utilization of highly-directional transmissions over wide bands in millimeter-wave (mmWave, $\approx$$30$\,GHz--$100$\,GHz, including the Ka band~\cite{pachler21updated}), sub-THz ($100$\,GHz--$300$\,GHz~\cite{De_Gaudenzi2022Open}), and even terahertz (THz, $300$\,GHz--$3$\,THz) frequencies is anticipated~\cite{meltem2021terahertz,aliaga2022joint}. Given the long propagation distances, both satellite-to-ground links (uplink and downlink) and inter-satellite links, also known as \emph{cross-links}, require the use of such directional high-rate connectivity. Additionally being actively considered for cross-links is the optical spectrum, which offers rates greater than even THz channels, and could complement them when the adequate channel conditions are given~\cite{Yahia2022HAPS}, but is greatly constrained by potential pointing losses due to extremely thin (i.e., laser-formed) beams~\cite{Kaushal2017optical, Jahid2022contemporary}. 

Utilizing mmWave radio for both access and cross-links offers the advantage of partial hardware reuse, resulting in simpler, lighter, and more cost-effective satellites. Although a compact, power and energy-efficient THz radio remains a long-term objective~\cite{Akyildiz2022terahertz}, this band is already being actively employed in satellite systems for sensing and imaging purposes~\cite{Roy2021Spaceborne, Brown2033Fundamentals}. Conversely, THz communications are progressing rapidly, supported by recent standardization initiatives~\cite{petrov2020ieee} and successful demonstrations of multi-kilometer-long THz links~\cite{Sen2022multi}, while the estimated link budget for satellite-to-airplane communication is deemed sufficient~\cite{Kokkoniemi2021channel}. Another benefit of mmWave and THz space-based cross-links is the reduced impact from terrestrial limitations, such as signal absorption by the atmosphere~\cite{Akyildiz2022terahertz}.

A significant hurdle in the advancement of mmWave and THz satellite communications pertains to the potential interference these solutions might introduce to the existing wireless systems, encompassing networks, radars, Earth exploration satellites, and more~\cite{Peng2022integrating, Perez2019Signal}. Despite many existing works in the field of satellite interference characterization and mitigation, which are reviewed in the following section, to the best of the author's knowledge, there is no study on the modeling of the interference among mmWave or THz satellite cross-links. As illustrated further in this article, interference between directional satellite cross-links needs to be carefully accounted for, when aiming for reliable data exchange within a single satellite constellation, as well as efficient co-existence of several LEO constellations in neighboring orbits.

\subsection{Related work}\label{subsec:related_work}
When analyzing satellite-based wireless systems, interference is one of the key factors considered in literature. Particularly, numerous studies focus on the \emph{interference between ground communication networks and satellite-based wireless systems}. Here, Sharma et al.~\cite{Sharma2012satellite} present one of the early attempts to capture the interference between a terrestrial base station and a satellite terminal, as well as review possible interference mitigation techniques. This work is followed by many further studies, including the recent ones by Zhang et al.~\cite{Zhang2022resource}, presenting the interference-aware resource allocation for C-band ($4$\,GHz--$8$\,GHz), Abdu et al.~\cite{Abdu2022Joint}, introducing a demand-aware resource allocation solution for interference-limited scenarios, and a survey of interference mitigating techniques by Peng et al.~\cite{Peng2022integrating}.

Among other methods, stochastic geometry (SG) has been identified as a particularly useful tool to model dense LEO constellations, where a terrestrial receiver is agnostic of the satellites' orbital parameters, as highlighted by Wang et al.~\cite{Wang2022Ultra}. Wang et al.~\cite{Wang2023Coverage} further use SG to analyze the coverage and data rate of an airplane receiver from a LEO network, and Okati et al.~\cite{Okati2020Downlink} utilize SG to model the total downlink performance of LEO constellations, among multiple other studies. Both works recognize the following: while SG offers a computationally efficient analysis tool, it typically yields approximate, time-averaged statistical results of performance in the presence of interference, ignoring any essential time-related dependencies the system may have.  

Going higher in frequency to mmWave and sub-THz spectrum, Xing and Rappaport~\cite{Rappaport2021terahertz} model potential interference between terrestrial and satellite-based systems operating over $100$\,GHz, followed by Kumar and Arnon~\cite{Kumar2022deriving} delivering an upper bound on ground-to-satellite channel capacity in W-band ($100$\,GHz--$110$\,GHz). Recently, dynamic spectrum sharing among terrestrial and satellite-based sub-THz wireless systems was experimentally demonstrated by Polese et al.~\cite{Polese2022dynamic}. While many of the analytical models discussed above consider a single satellite, there have been several studies explicitly capturing the interference produced by a group of satellites, including but not limited to~\cite{Tonkin2018newspace},~\cite{Braun2019should}, and~\cite{Lin2021method}.  Still, the key focus of these works is on the \emph{interference between the satellite-based systems and terrestrial networks}.

As the total volume of satellites keeps rapidly growing, the potential interference among the satellites themselves becomes an issue. Particularly, Fortes and Sampaio-Neto studied the interference between a LEO satellite and a Medium Earth orbit (MEO) satellite back in 2003~\cite{Fortes2003Impact}. Since then, dozens of studies have been presented for \emph{cross-orbit interference} in S-band ($2$\,GHz--$4$\,GHz)~\cite{Mendoza2017spectrum} and Ka-band ($27$\,GHz--$40$\,GHz)~\cite{Sharma2016inline,Wang2018coexistance}. Some latest studies, including~\cite{Leyva2021inter} and~\cite{Leyva2021inter2}, focus on LEO-LEO satellite interference at $2.4$\,GHz and $20$\,GHz.


Importantly, to the best of the authors' knowledge, while the density of the satellite communication networks passed already several tens of satellites per orbit in deployed or announced constellations~\cite{spaceXGen1,kuiper} and keeps growing rapildy~\cite{spaceXGen2}, \emph{there is no comprehensive mathematical model that studies the interference among LEO satellite cross-links operating at mmWave and THz frequencies}. Most of the prior works are limited to either: (i)~ground-to-satellite interference (\cite{Sharma2012satellite,Zhang2022resource,Peng2022integrating,Rappaport2021terahertz,Kumar2022deriving,Polese2022dynamic,Tonkin2018newspace,Braun2019should,Lin2021method} among others), (ii)~lower frequencies (\cite{Leyva2021inter,Leyva2021inter2} among a few others), or (iii)~simulation-based studies (\cite{Mendoza2017spectrum,Sharma2016inline,Wang2018coexistance} among others). Meanwhile, existing interference models for mmWave and THz networks (\cite{Venugopal2016device},~\cite{Jornet2011low}, and~\cite{Petrov2017interference} among others) are not directly applicable, as they model terrestrial setups and do not account for the specific satellite systems' scenario geometry and other important peculiarities of cross-link operation. By expanding our preliminary work~\cite{sergi_ICC_2023}, we aim to address this gap in the present article.

\subsection{Claims and contributions}\label{subsec:claims_contrib}
Concerned by the lack of comprehensive frameworks that capture the realistic impact of directional cross-link interference in prospective mmWave/THz LEO satellite communication systems, \emph{in this paper}, we present a mathematical framework and an extensive numerical study to model directional interference among mmWave/THz cross-links and its impact on the key performance indicators (KPIs). The presented models, results, and conclusions should facilitate further development and evaluation of possible constellation designs for reliable high-rate mmWave and THz LEO satellite networks, as well as the seamless co-existence of several constellations deployed in the proximity of each other.

The contributions of this work are summarized as follows:
\begin{compactitem}
 \item \emph{Novel analytical methodology}: An elaborate mathematical framework is developed to model directional interference among cross-links in mmWave/THz satellite communication networks. All the key configurations are considered, including the interference coming from the same orbit, a shifted orbit, and a co-planar orbit. The framework takes into account all the major static and time-variant orbital parameters of the involved LEO satellites, their mutual orientation and mutual mobility, as well as the essential radio link parameters.

 \item \emph{In-depth numerical study}: A thorough investigation is performed using the developed methodology of the directional mmWave/THz LEO satellite communication system in the presence of cross-link interference. The study highlights time-variant, statistical, and time-averaged levels of interference power, as well as the impact of cross-link interference on other KPIs, including the signal-to-interference ratio (SIR), signal-to-interference-plus-noise ratio (SINR), and link capacity. The study particularly highlights the cases, where cross-link interference can be neglected and the setups where it should not be ignored. The latter characterises the specific deployment configurations that may benefit from interference mitigation techniques, including but not limited to channelization and interference-aware channel access.

 \item \emph{Comprehensive simulation study}: The results delivered with the mathematical framework are further cross-verified by our in-house built simulator for satellite wireless systems. The developed simulation framework captures the design and mobility of the realistic constellation of LEO satellites. The presented analytical results demonstrate a close match to those delivered with the simulation framework, thus confirming the accuracy of our delivered models and the trustfulness of the obtained numerical conclusions.
\end{compactitem}

The rest of the paper is organized as follows. In Sec.~\ref{sec:system_model} we review the main parameters necessary to characterize the satellite's position and movement, we introduce the antenna and network models for our study and we also outline the selected deployment scenarios that are analyzed. The mathematical models of interference, SIR, SINR and channel capacity are derived in Sec.~\ref{sec:interference_analysis}. We evaluate numerical and analytical results, as well as provide a detailed description of the simulator in Sec.~\ref{sec:results}. Finally, key observations, conclusions, and future work are outlined in Sec.~\ref{sec:conclusions}. 

\begin{figure}[!b]
    \centering
    \subfigure[High-level deployment illustration of three shifted orbits]
    {
        \includegraphics[width=0.48\textwidth]{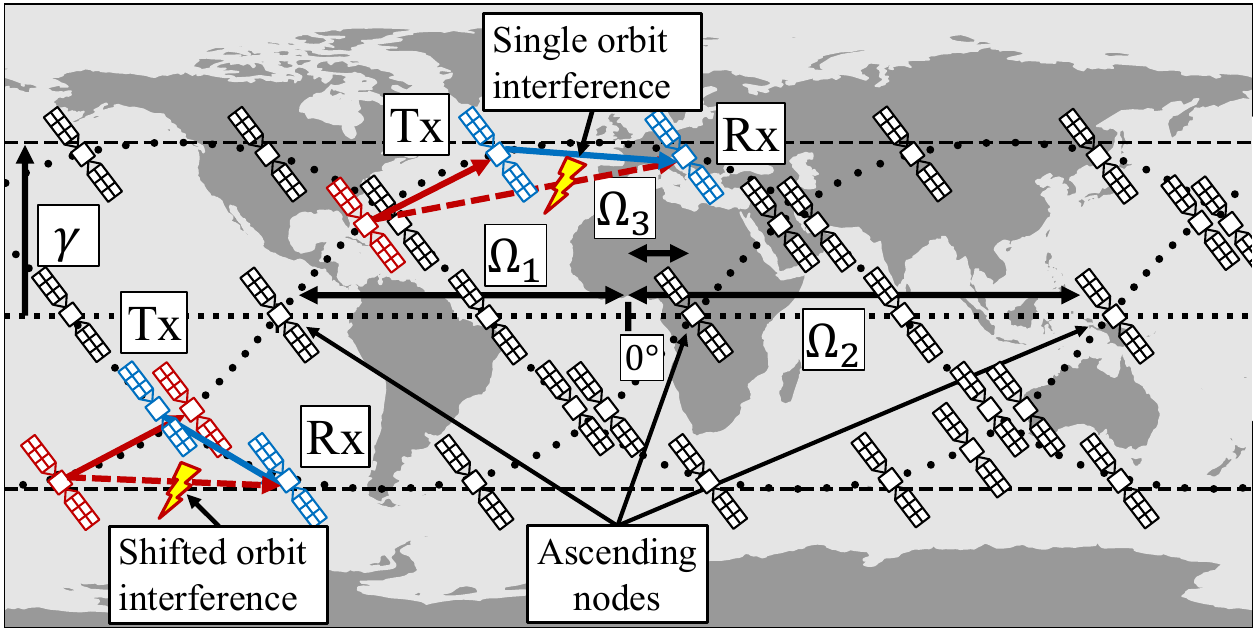}
        \label{fig:orb_2d}
    }
    \subfigure[Single and co-planar LEO interference]
    {
        \includegraphics[width=0.46\textwidth]{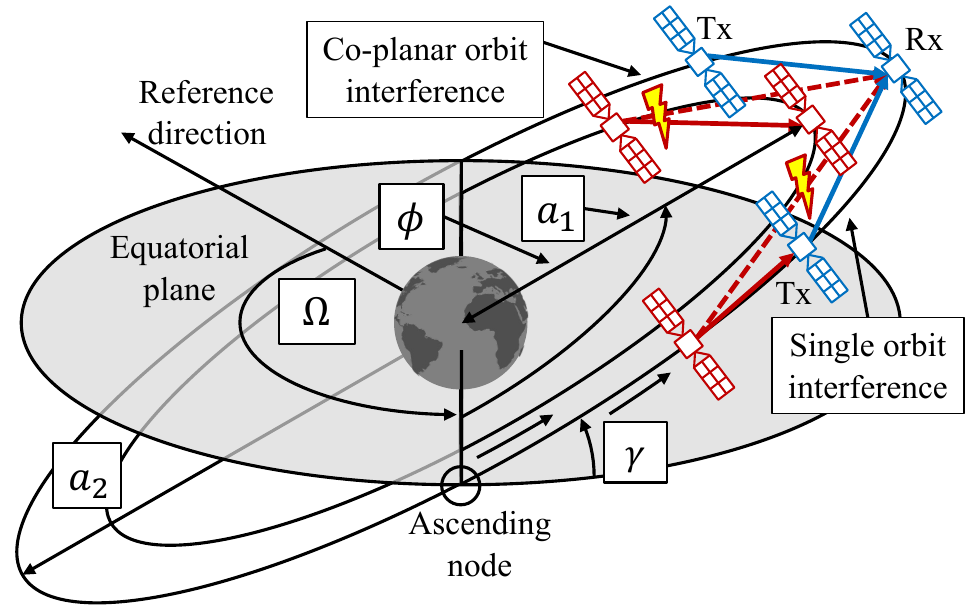}
        \label{fig:single_coplanar}
    }
    \subfigure[Shifted and shifted co-planar LEO interference]
    {
        \includegraphics[width=0.46\textwidth]{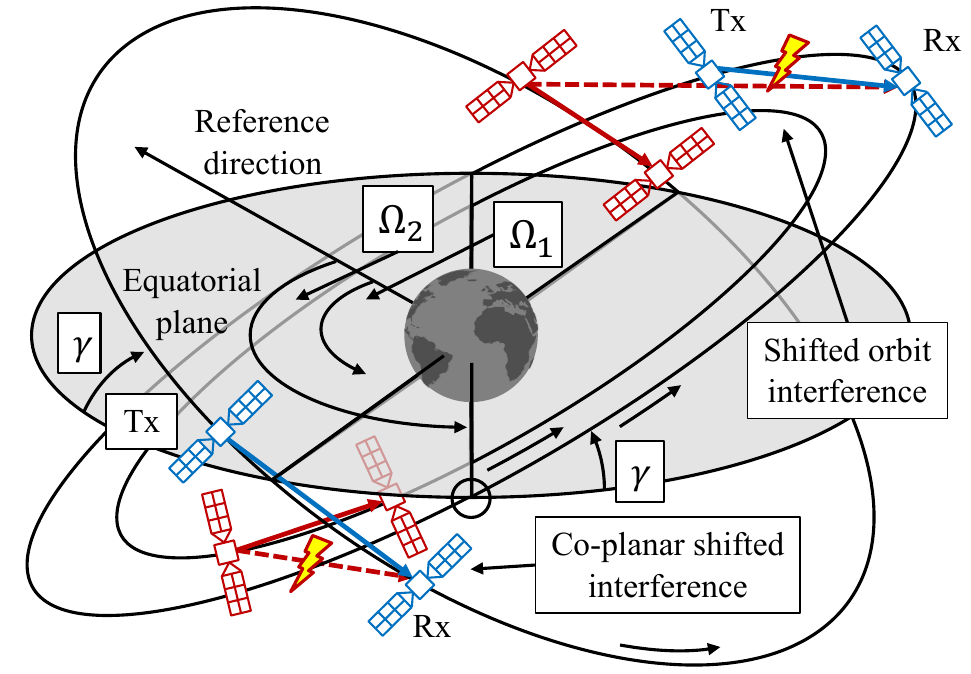}
        \label{fig:shifted_shifted_coplanar}
    }
    \captionsetup{justification=centering}
    \vspace{-2mm}
    \caption{Interference modeling for LEO  mmWave/THz directional cross-link communications.}
    \vspace{-8mm}
    \label{fig:system_model}
\end{figure}


\section{System Model}
\label{sec:system_model}

In this section, the main assumptions for our model are introduced. Table~\ref{tab:notation} summarizes the key notation.

\begin{table}[b!]
\footnotesize
\centering
\vspace{-6mm}
\caption{Notation used in the interference analysis}
\begin{tabular}{p{0.16\columnwidth}p{0.71\columnwidth}}
\hline
\textbf{Notation}&\textbf{Description} \\
\hline
&\textbf{\textit{Orbital parameters}} \\
\hline
$R_{\textrm{E}}$ & \textcolor{black}{Earth radius}\\
$a$ & Semi-major axis \\
$\gamma$ & Inclination \\
$\phi$ & True anomaly\\
$\Omega$ & Right Ascension of the Ascending Node (RAAN)\\
$h$, $h_\text{C}$, $h_\text{S}$ & Orbital altitude (single, co-planar, shifted)\\
$N$, $N_\text{C}$, $N_\text{S}$ & Number of satellites per orbit (single, co-planar, shifted)\\
$N_1$ & Number of interfering satellites in the orbit of interest\\
$N*$ & \textcolor{black}{Optimal number of satellites to maximize capacity}\\
$\beta$ & Relative angular offset\\
$\Delta\beta$ & Difference in relative angular offset between two orbits\\
$\Delta\beta_{\text{max}}$ & Maximum relative angular offset before redundancy\\
$i$, $j$ & Satellite indexes in the current orbit and the secondary orbit\\
\hline
&\textbf{\textit{Radio and propagation parameters}} \\
\hline
$P_{\text{Tx}}$, $P_{\text{Rx}}$ & Transmit and received power\\
$G$ & Antenna gain\\
$B$ & Channel bandwidth\\
$\alpha$ & Antenna radiation pattern angle (beamwidth) \\
$f_\text{C}$ & Carrier frequency \\
$\lambda$ & Carrier wavelength\\
$k$ & Boltzmann's constant\\
$T$ & System temperature\\
\hline
&\textbf{\textit{Performance metrics}} \\
\hline
$E[I]$ & Average interference power at the lawful receiver\\
$\Gamma$ & \textcolor{black}{Signal-to-interference ratio}\\
$S$ & \textcolor{black}{Signal-to-interference-plus-noise ratio}\\
$C$ & Channel capacity\\
\hline
\end{tabular}
\label{tab:notation}
\end{table}

\subsection{Orbital parameters}
Figure~\ref{fig:system_model} illustrates the considered scenarios. In this study, we particularly focus on the orbits commonly used for LEO satellites. These circular orbits are primarily determined by four essential parameters, namely, the \textit{semi-major axis} $a$, \textit{inclination} $\gamma$, \textit{true anomaly} $\phi$, and \textit{right ascension of the ascending node} (RAAN), $\Omega$, as in Figs.~\ref{fig:single_coplanar},~\ref{fig:shifted_shifted_coplanar}, and explained below. The semi-major axis $a$ is the distance from the center of the Earth to the satellite's orbit, which is determined by the orbit altitude above sea level, $h$, and the Earth's radius, $R_{\textrm{E}}$, as $a=R_{\textrm{E}}+h$. The inclination $i$ is the angle between the orbital plane and the Earth's equatorial plane. The true anomaly $\phi$ is the angle between the satellite's position and the \textit{ascending node}, which is the point where the orbit intersects the Earth's equator when moving from south to north. The final parameter, the RAAN $\Omega$, determines the angular position of the ascending node with respect to a reference direction, which is typically determined by the vernal equinox. Satellites on each orbit are evenly spaced from each other.

\subsection{Selected scenarios}
The presented study aims to comprehend the interference levels, sources of interference, and interference patterns over time in directional satellite communications by breaking down the problem into mathematically tractable scenarios. We have identified three practical sub-scenarios of varying complexity that serve as the foundation for the upcoming satellite mega-constellations currently being deployed or designed. Our approach allows us to analyze each sub-scenario in detail and gain insights into the overall interference problem in the next generation of satellite mega-constellations. These identified sub-scenarios, illustrated in Fig.~\ref{fig:system_model} are:
\begin{compactenum}
    \item \emph{Single orbit}: The simplest option, one orbit with evenly distributed satellites. Interference comes from the satellites located in the same orbit.
    \item \emph{Co-planar orbits}: A more sophisticated case, an additional orbit in the same orbital plane but at a higher altitude. This case is of particular importance to the co-existence of multiple constellations with similar parameters to offer competitive coverage but different altitudes.
    \item \emph{Shifted orbits}: The most complex sub-scenario analyzed. Multiple shifted orbits are typically combined to form a grid pattern between various latitudes, giving rise to the distinctive satellite ``shells" of what are known as Walker constellations~\cite{Leyva2022ngso}.
\end{compactenum}
We describe the three scenarios in more detail in Sec.~\ref{sec:problem_formulation}.

\subsection{Propagation, antenna and routing assumptions}
Our research focuses on low Earth orbit (LEO) satellites that are positioned at altitudes ranging from $500$\,km to $2000$\,km, where atmospheric absorption has a negligible effect compared to spreading~\cite{Kokkoniemi2021channel}. We assume that all the satellites transmit at the same power level, $P_{Tx}$, and utilize equivalent antenna systems for transmission and reception\footnote{In a homogeneous satellite network performing half-duplex or full-duplex wireless communications over cross-links, it is most realistic to assume that the same or identical antenna modules are used for both transmission and reception. The developed framework can be extended further to model different antenna gains ($G_{\text{Tx}} \neq G_{\text{Rx}}$) and/or different radiation patterns with e.g., different beamwidths at the transmitter and receiver sides.}. Therefore, the received power, $P_{Rx}$, depends primarily on the transmit power, antenna gains, and spreading loss, which can be expressed as $P_{Rx} = \frac{P_{Tx}G^2}{(4\pi d / \lambda)^2}$. Here, $P_{Tx}$, $G$, $d$, and $\lambda$ represent the transmit power, antenna gain, separation distance between the transmitter and the receiver, and signal wavelength, respectively.


To analyze interference in a satellite constellation, certain assumptions need to be made about the routing strategy used. In our study, we assume that satellites transmit to their immediate neighbors in the same orbit whenever possible. This approach is feasible for the modeled scenarios, as global coverage can be achieved with first-neighbor links, avoiding the need for excess output power to reach farther neighbors~\cite{Royster2023network}. While inter-orbit cross-links may also be needed, our preliminary study of the scenario geometries and dynamics revealed that their impact on the total interference is of secondary importance compared to direct in-orbit links. Therefore, these types of links are excluded from our first-order study for the sake of tractability of the resulting mathematical framework and clarity of the key numerical results. However, the model elements we developed can be recombined to account for different topologies and routing protocols. We also assume that all communicating nodes use an analytical cone-shaped antenna radiation pattern with a main lobe gain given by $G=2/(1-\cos(\alpha/2))$, where $\alpha$ is the pattern directivity angle/beamwidth~\cite{Petrov2017interference}. The delivered framework can be further tailored to other radiation patterns, both with or without strong side lobes.

\subsection{Metrics of interest}\label{sec:metrics_of_interest}
The main emphasis of our study is to examine the average power of interference that occurs at the intended receiving node, which is denoted as $E[I]$. Our analysis takes into account the combined impact of all the feasible satellite sources that could potentially cause interference. Additionally, we compare the estimated interference value with both the strength of the received signal and the power of the Johnson-Nyquist thermal noise, using the signal-to-interference ratio (SIR) and the signal-to-interference-plus-noise ratio (SINR), respectively. Lastly, we examine the overall channel capacity, $C$, calculated as $C = B \log_2(1+\text{SINR})$.

\section{Interference Analysis}
\label{sec:interference_analysis}
\label{sec:problem_formulation}

\subsection{General approach}
Despite their deterministic nature marked by orbital physics, the geometries of LEO satellite constellations feature notably complex and, importantly, time-dependent mutual location and orientation of the individual satellites with non-trivial interactions among each other. Therefore, for tractability, we first have to decompose the overall setup of a typical Walker constellation~\cite{walker1984satellite} into reduced sub-parts of the overall constellation and analyze those separately.

We characterize the interference originating from satellites in the same orbit as the target receiver in Section~\ref{sec:single}, while analyzing the theoretical performance limits in Section~\ref{sec:single_limits}. Next, we investigate the interference caused by satellites positioned in a co-planar orbit in Section~\ref{sec:coplanar}. Subsequently, in Section~\ref{sec:shifted} we present the model and analysis for interference arising from satellites in a shifted orbit, while in Section~\ref{sec:shifted_coplanar} we extend this analysis to a shifted orbit at a different altitude to a co-planar shifted orbit. Each section collectively completes the general setup depicted in Figure~\ref{fig:system_model}.

\subsection{Modeling interference from the same orbit}
\label{sec:single}
\begin{figure}[h]
  \centering
  \vspace{-3mm}
  \includegraphics[width=0.5\textwidth]{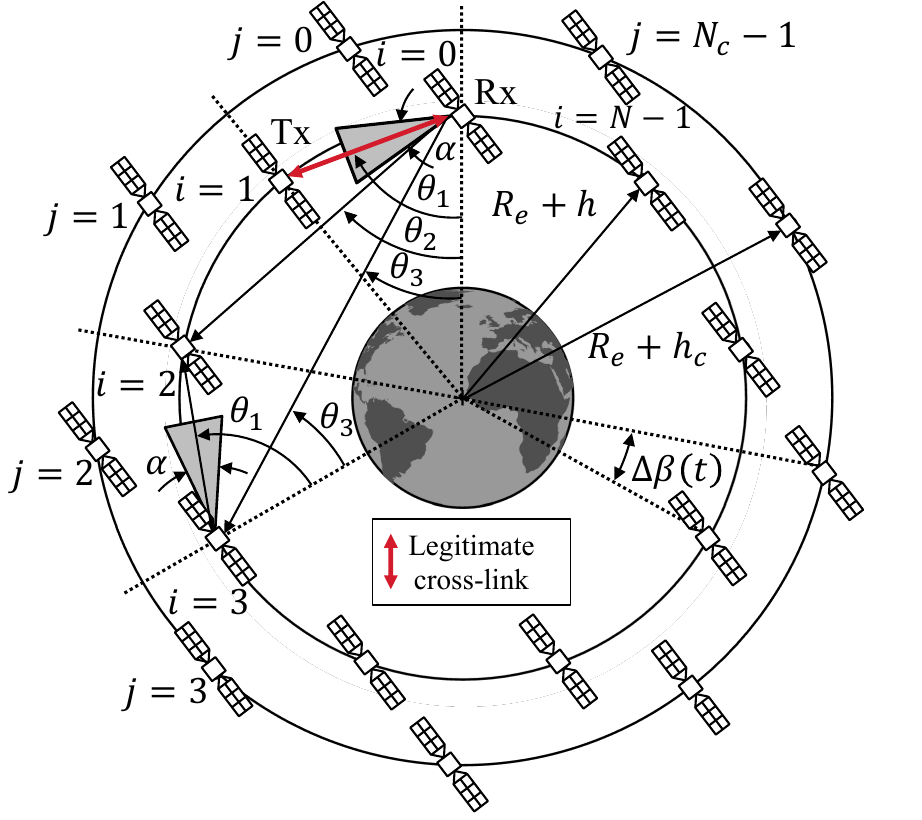}
  \vspace{-3mm}
  \caption{Modeling single and co-planar orbits interference.}
  \label{fig:single_orbit_geometry}
  \vspace{-3mm}
\end{figure}

Focusing on the interference coming from the same orbit as the target receiver, the setup geometry is illustrated in Fig.~\ref{fig:single_orbit_geometry}, along with a co-planar orbit setup analyzed in the following subsection. Index $i$ is used to number the satellites in the orbit of interest ($i \in [0; N-1]$), where $N$ is the total number of satellites, and the link between satellites $i=1$ and $i=0$ is considered as the link of interest (the signals coming from other satellites $i\neq\{0,1\}$ that are captured at the receiver, $i=0$, are considered as interference). Since the satellites of a single orbit reside in the same orbital plane, a 2D simplification of the problem can be considered when analyzing the interference from that same orbit $I_1$.

In general, a transmitting satellite will cause interference to the target receiver if and only if the three following conditions are jointly satisfied: (i)~the interfering satellite has line-of-sight with the receiver, i.e., it is not blocked by the Earth, (ii)~the interfering satellite appears within the receiver beam, and (iii)~the receiver appears within the interfering satellite beam as well. 
First, the interfering satellite is in line of sight with the receiver if:
\begin{equation}\label{eq:cond1}
    \theta_i > \frac{\pi}{2} - \arccos\left(\frac{R_{\textrm{E}}}{R_{\textrm{E}} + h}\right),\ i\in [2, N].
  \end{equation}

Second, given the radial symmetry presented in the single orbit scenario depicted in Fig.~\ref{fig:single_orbit_geometry}, the second and third conditions are satisfied if:
\begin{equation}\label{eq:cond23}
    \theta_i > \theta_1 - \frac{\alpha}{2},\ i\in [2, N].
  \end{equation}

Considering that the satellites are evenly distributed in orbit, an expression for the angle $\theta_i$ can be obtained using trigonometry as:
\begin{equation}\label{eq:theta_i}
  \theta_i = \frac{\pi}{2} - i\frac{\pi}{N},\ i\in [1, N].
\end{equation}

Thus, substituting (\ref{eq:theta_i}) into (\ref{eq:cond1}) and (\ref{eq:cond23}) results in a mathematical expression of the set of satellites causing interference to the target receiver:
\begin{equation}
  i < \frac{N}{\pi} \arccos\left(\frac{R_{\textrm{E}}}{R_{\textrm{E}} + h}\right)\enspace{}\land\enspace{}i < 1 + \frac{N}{2\pi}\alpha,\ i\in [2, N].
\end{equation}

In other words, the total number of interfering satellites, $N_{1}$, is set by the minimum of these two thresholds. Remembering that satellite $i=1$ is the transmitter in the link of interest and not an interfering satellite, we obtain:
\begin{equation}\label{eq:n_1_single}
  N_1 = \left \lfloor \min\left(\frac{N}{\pi} \arccos\left(\frac{R_{\textrm{E}}}{R_{\textrm{E}} + h}\right), 1 + \frac{N}{2\pi}\alpha\right) \right \rfloor - 1.
\end{equation}

To calculate the signal power captured by the receiver,  it is also necessary to define the distance $d_i$ between the receiver and any other satellite $i$. For satellites in the same orbit, this distance will remain constant through time since satellites in the same orbit travel at the same speed, i.e., their relative speed is zero. Applying the Law of cosines in Fig.~\ref{fig:single_orbit_geometry} leads to:
\begin{equation}\label{eq:single_orb_distance}
  d_i^2 = 2\left(R_{\textrm{E}}+h\right)^2\left(1-\cos\frac{2\pi}{N}i \right).
\end{equation}

From here, an expression for the average interference power for the single orbit scenario, $E[I_1]$, can be derived by adding all the individual contributions of the satellites that satisfy the three initial conditions. Combining \eqref{eq:single_orb_distance} and \eqref{eq:n_1_single} with the propagation model from Sec.~\ref{sec:system_model}, we obtain:
\begin{equation}\label{eq:i1}
  \hspace{-2mm}E[I_1] = \sum_{i=2}^{N_1+1} \frac{\lambda^2 P_{Tx}}{8\pi^2\left(1-\cos\frac{\alpha}{2}\right)^2\left(R_{\textrm{E}}+h\right)^2\left(1-\cos \frac{2\pi}{N}i\right)}.
\end{equation}

By calculating the distance between the transmitter and the receiver, $d_1$ in \eqref{eq:single_orb_distance}, the first metric of interest for the single orbit scenario, that is, the SIR, can be analytically derived:
\begin{equation}\label{eq:sir1}
  \Gamma_1 = \frac{\frac{1}{\left(1-\cos\frac{2\pi}{N} \right)} }{ \sum_{i=2}^{N_1+1} \frac{1}{\left(1-\cos\frac{2\pi}{N}i \right)}}.
\end{equation}

It is important to note that many input parameters, such as the Tx power or the orbital altitude, cancel in the numerator and denominator, so the average SIR depends primarily on the number of satellites in orbit. A further elaborated analysis is provided in Sec.~\ref{sec:results}. 

An expression for the SINR can thus be derived by considering the Johnson-Nyquist noise power $P_N=kTB$, where $T$ is the system's temperature, $B$ is the bandwidth of the communications system, and $k$ is the Boltzmann constant:
\vspace{-2mm}
\begin{equation}\label{eq:snr1}
    S_1 = \frac{\frac{\lambda^2P_{Tx}}{8\pi^2(1-cos\frac{\alpha}{2})^2(R_{\textrm{E}}+h)^2(1-cos\frac{2\pi}{N})}}{\sum_{i=2}^{N_1+1}\frac{\lambda^2P_{Tx}}{8\pi^2(1-cos\frac{\alpha}{2})^2(R_{\textrm{E}}+h)^2(1-cos\frac{2\pi}{N}i)}+kTB}
\end{equation}

\subsection{Single orbit: Theoretical performance limits}\label{sec:single_limits}
As in any communication system, a trivial solution to maximize the $\text{SNR}$ would be to reduce the distance between the transmitter and the receiver, which in the case of an intra-orbit cross-link would be directly related to deploying more evenly spaced satellites in that orbit. Here, an interesting trade-off arises: on the one hand, considering more satellites will indeed increase the $\text{SNR}$ by bringing the transmitter closer to the receiver, while, on the other hand, more satellites will increase the expected interference, thus degrading the $\text{SIR}$ and the overall performance of the link. Formalizing this trade-off analytically:
\begin{equation}\label{eq:N_lim}
    \begin{split}
        &\lim_{N\to\infty} S_1(N) = \lim_{N\to\infty} \Gamma_1(N) =  \\
        & \lim_{N\to\infty} \frac{1}{\left(1-\cos\frac{2\pi}{N} \right)} / \sum_{i=2}^{N_1+1} \frac{1}{\left(1-\cos\frac{2\pi}{N}i \right)} =  \\
        & = 1 / \sum_{i=2}^{\infty}\frac{1}{i^2} = \frac{1}{\frac{\pi^6}{6}-1} \approx 1.55 = 1.9 \text{dB}
    \end{split}
\end{equation}
where we have utilized the Taylor series expansion $\cos(x)\approx1-\frac{x^2}{2}$ and the infinite series $\sum^\infty_{i=2}\frac{1}{i^2}=\frac{\pi^6}{6}-1$. This result implies that the cross-link channel capacity can not be arbitrarily increased by adding more satellites to the constellation, but rather there is a limit for which the improvement in $\text{SNR}$ is compensated by the increase in received interfering power, even if the received signal strength is well above the noise floor (i.e. low $\text{SIR}$ limits channel capacity even if the $\text{SNR}$ is high). This is an important theoretical result from our work that is further elaborated on in Sec.~\ref{sec:results}.

The natural question then is \textit{what is the maximum number of satellites per orbit after which the cross-link performance starts degrading due to the interference them for a given orbital altitude $h$ and antenna beamwidth $\alpha$?} Although the number of satellites to be deployed is typically found through optimizing the global coverage and capacity of terrestrial receivers, with the presented framework we can derive this upper bound beyond which adding more satellites is detrimental to cross-link performance. This bound represents an important constraint to consider in multi-parameter optimization when designing next-generation satellite communication networks with high-rate directional cross-links. Provided that the system noise power $P_N$ is low enough to provide an $\text{SNR}$ higher than the limit of 1.9~dB, the maximum $\text{SINR}$ will be obtained with the maximum number of satellites that can grant no unlawful transmissions reaching the receiver, that is $N_1=0$. Further analyzing \eqref{eq:n_1_single}, we can derive this optimal number of satellites to deploy, $N^*$, by solving the inequality $N_1<1$, which taking into account the two thresholds given by the interference conditions results into:
\begin{equation}\label{eq:N_optimal}
    N^* = \max\left(\frac{2\pi }{ \alpha\ } ,\  \frac{2\pi }{ \arccos\left(\frac{R_{\textrm{E}}}{R_{\textrm{E}}+h}\right)}\right)
\end{equation}
This result is further explored in Sec.~\ref{sec:results} as well.

\subsection{Co-planar orbits}
\label{sec:coplanar}
\begin{figure}[!t]
  \centering
  \includegraphics[width=0.45\textwidth]{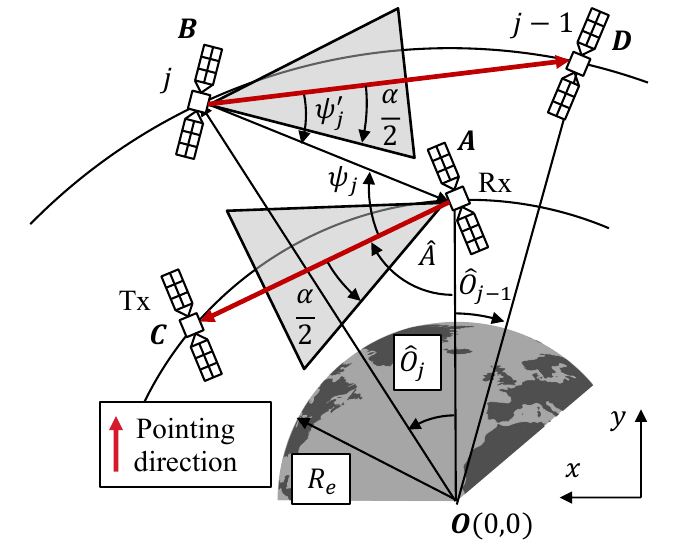}
  \caption{Geometry of a co-planar orbit interference setup.}
  \label{fig:coplanar_orbits_zoom}
  \vspace{-6mm}
\end{figure}

A co-planar orbit setup is also depicted in Fig.~\ref{fig:single_orbit_geometry}. This scenario illustrates a situation in which two separate satellite service providers have deployed part of their constellations in the same orbital plane, at different altitudes, which is the reason why we only consider links between satellites in the same orbit. Given the current satellite communications landscape, this scenario is likely to occur, since multiple LEO service providers might be interested in giving coverage to the same areas. A closer look at the geometry of the setup is given in Fig.~\ref{fig:coplanar_orbits_zoom}, with $N$ satellites in the lower orbit and $N_{c}$ satellites in the higher orbit. A pair of Cartesian axes $x$ and $y$ is introduced to simplify the mathematical formulation using vector notation. 

In this case, as the two orbits have different altitudes, they also have different orbital speeds. For this reason, the time-dependent parameter $\Delta\beta(t)$, referred to as \emph{relative angular offset}, is introduced. This parameter captures the periodicity of the interference patterns over time due to the difference in orbital periods, determined by Kepler's third law~\cite{Maral2009satellite}. An accurate description of the relation between $\Delta\beta(t)$ and the duration of the interference patterns in a two-orbit setup (not necessarily co-planar) is provided in Appendix~\ref{app:rel_angular_offset}. The time argument is omitted from the formulation going forward for the sake of clarity. 


Again, the three conditions from the previous subsection have to be met for the satellite $j$ to interfere with the receiver. However, as for co-planar orbits $h \neq h_\text{C}$, condition~2 and condition~3 are now fulfilled separately. Following Fig.~\ref{fig:coplanar_orbits_zoom} formulation,  Condition~1 is satisfied if:
\begin{equation}
  \psi_j + \hat{A} > \frac{\pi}{2} - \arccos\left(\frac{R_{\textrm{E}}}{R_{\textrm{E}} + h}\right).
\end{equation}

In their turn, mutual antenna alignment Conditions 2 and 3 are satisfied if $|\psi_j| < \frac{\alpha}{2}$ and $| \psi_j' | < \frac{\alpha}{2}$, respectively.

Using vector notation, we then derive the angle $\psi_j$ as:
\vspace{-1mm}
\begin{equation}\label{eq:psi_j}
  \psi_j = \arctan\left(\| \vec{AC} \times \vec{AB} \| / \langle \vec{AC}, \vec{AB} \rangle \right) \in \{-\pi, \pi\}
\end{equation}
\vspace{-1mm}
where
\begin{align}
  &\vec{A} = (0 ,\quad{}\hspace{-1mm} R_{\textrm{E}} + h),\ \vec{B} = (R_{\textrm{E}} + h_\text{C}) (\sin\hat{O}_j ,\quad{}\hspace{-1mm} \cos\hat{O}_j) \label{vecB},\\
  &\vec{AB} = \vec{B}\hspace{-0.5mm}-\hspace{-0.5mm}\vec{A} = ((R_{\textrm{E}} + h_\text{C})\sin\hat{O}_j ,\quad{}\hspace{-1.5mm} (h_\text{C}-h)\cos\hat{O}_j),\\
  &\vec{AC} = (\sin\hat{A} ,\quad{} -\cos\hat{A}),\quad{}\hat{O}_j = \Delta\beta(t) + 2\pi j / N_\text{C}.
\end{align}

With vector notation, angle $\hat{A}$ corresponds to the angle $\theta_1$ from the single orbit scenario as $\hat{A} = \frac{\pi}{2} - \frac{\pi}{N}$. Therefore, (\ref{eq:psi_j}) can be expanded into
\vspace{-1mm}
\begin{equation}
    \hspace{-1mm}\psi_j\hspace{-0.7mm}= \hspace{-0.7mm}\arctan\hspace{-1mm}\left(\hspace{-1.5mm}\frac{(h_\text{C}\hspace{-1mm}-\hspace{-1mm}h)\hspace{-0.7mm}\cos\hspace{-0.7mm}\frac{\pi}{N}\cos\hspace{-0.7mm}\hat{O}_j\hspace{-1mm}+\hspace{-1mm}(R_{\textrm{E}}\hspace{-1mm}+\hspace{-1mm}h_\text{C})\sin\hspace{-0.7mm}\frac{\pi}{N}\sin\hspace{-0.7mm}\hat{O}_j}{(R_{\textrm{E}}\hspace{-1mm}+\hspace{-1mm}h_\text{C})\hspace{-0.7mm}\cos\hspace{-0.7mm}\frac{\pi}{N}\sin\hspace{-0.7mm}\hat{O}_j\hspace{-1mm}-\hspace{-1mm}(h_\text{C}\hspace{-1mm}-\hspace{-1mm}h)\sin\hspace{-0.7mm}\frac{\pi}{N}\cos\hspace{-0.7mm}\hat{O}_j}\hspace{-1.5mm}\right)\hspace{-1mm}.\hspace{-1mm}
\end{equation}

The other important angle, $\psi_j'$, which is only relevant in absolute value, can therefore be obtained as:
\vspace{-1mm}
\begin{equation}
  \psi_j' = \arccos\left( \langle \vec{BD}, \vec{BA} \rangle / \left[ \| \vec{BD} \| \cdot \| \vec{BA}\| \right] \right),
\end{equation}
\vspace{-1mm}
where
\vspace{-1mm}
\begin{align}
  &\vec{BD} = \vec{D} - \vec{B},\quad{}\quad{}\vec{BA} = \vec{A} - \vec{B},\\
  &\vec{D} = (R_{\textrm{E}} + h_\text{C})(\sin\hat{O}_{j-1},\quad{} \cos\hat{O}_{j-1}),\\
  &\hat{O}_{j-1} = \Delta\beta(t) + 2\pi/N_\text{C} ((j-1)\mod (N_\text{C}-1)).
\end{align}

After deriving the main angles, $\psi_j$ and $\psi_j'$, the set of interfering satellites $C = C1 \cap C2 \cap C3$, which satisfy Conditions 1, 2, and 3, respectively, can be defined as:
\begin{align}
  C_1 &= \{j \in [0,N_\text{C}-1]\ |\ \psi_j > \frac{\pi}{N} -\arccos\left(\frac{R_{\textrm{E}}}{R_{\textrm{E}} + h}\right)\} ,\nonumber\\
  C_2 &= \{j \in [0,N_\text{C}-1]\ |\ |\psi_j| < \frac{\alpha}{2} \},\\
  C_3 &= \{j \in [0,N_\text{C}-1]\ |\ |\psi_j'| < \frac{\alpha}{2} \}.\nonumber
\end{align}

The distance between the possible interferer $j$ and the target receiver is now calculated. This distance corresponds to $\| \vec{AB} \|$, according to the analysis above and the geometry in Fig.~\ref{fig:coplanar_orbits_zoom}, thus:
\begin{equation}
  d_j^2 = (R_{\textrm{E}}+h_\text{C})^2\sin^2\hat{O}_j+(h_\text{C}-h)^2\cos^2\hat{O}_j.
\end{equation}

Then, the average interference power received from the higher orbit, $E[I_2]$, can be computed as follows:
\begin{align}\label{eq:i2}
  E[I_{2}] = &  \sum_{j \in C_1\cap C_2 \cap C_3}\lambda^2 P_{Tx} / [4\pi^2 \left(1-\cos\frac{\alpha}{2}\right)^2 \nonumber\\
  &((R_{\textrm{E}}+h_\text{C})^2 \sin^2\hat{O}_j) + (h_\text{C}-h)^2\cos^2\hat{O}_j)],
\end{align}

By solving the following equation, we can also find the minimum co-planar orbit altitude that ensures interference isolation between orbits, $h_\text{C}^*$:
\begin{align}\label{eq:h_c_star}
  &\hspace{-3mm}E[I_{2}] = \sum_{j \in C_1\cap C_2 \cap C_3} \lambda^2 P_{Tx} / [4\pi^2 \left(1-\cos\frac{\alpha}{2}\right)^2 \nonumber \\
   & \hspace{-3mm} ((R_{\textrm{E}}+h_\text{C}^*)^2 \sin^2\hat{O}_j) + (h_\text{C}^*-h)^2\cos^2\hat{O}_j)] = 0, \forall \hat{O}_j(t)
\end{align}

If we now consider the total interference, both from the lower and higher orbits, the signal-to-interference ratio for a link in the lower orbit, $\Gamma_2$, can be computed as:
\begin{align}\label{eq:sir2}
    &\hspace{-3mm} \Gamma_2 = \frac{P_{\text{Rx}}}{E[I_{1}]+E[I_{2}]} = \nonumber \\
    &\hspace{-3mm}=\frac{\frac{1}{2\left(1-\cos\frac{2\pi}{N} \right)} }{  \sum\limits_{i=2}^{N_1+1} \frac{1}{2\left(1-\cos\frac{2\pi}{N}i \right)} + \hspace{-2.5mm}\sum\limits_{j \in C_1\cap C_2 \cap C_3} \hspace{-1mm} \frac{1}{\sin^2\hat{O}_j + \frac{(h_\text{C}-h)^2}{(R_{\textrm{E}}+h_\text{C})^2} \cos\hat{O}_j}}.
\end{align}

Analogously to section~\ref{sec:single}, expressions for $S_2$ can be obtained when considering the thermal noise at the receiver: 
\begin{equation}
    S_2 = \frac{P_{\text{Rx}}}{E[I_{1}]+E[I_{2}]+kTB}
\end{equation}
where expressions \eqref{eq:i1} and \eqref{eq:i2} are used for the calculation of $E[I_{1}]$ and $E[I_{2}]$ respectively.

An equivalent analysis can also be formulated to study the metrics of interest for a cross-link in the higher orbit. The resulting expressions for $E[I_{2c}]$, $\Gamma_{2\text{C}}$ and $S_{2\text{C}}$ are, respectively:
\begin{align}\label{eq:i2c}
  E[I_{2c}] &= \sum_{i \in C_1\cap C_2 \cap C_3}  \lambda^2 P_{Tx} / [4\pi^2 \left(1-\cos\frac{\alpha}{2}\right)^2 \nonumber\\
   &((R_{\textrm{E}}+h_\text{C})^2 \sin^2\hat{O}_i) + (h_\text{C}-h)^2\cos^2\hat{O}_i)],
\end{align}
\vspace{-4mm}
\begin{align}\label{eq:sir2c}
    \Gamma_{2\text{C}} &= \frac{P_{\text{Rxc}}}{E[I_{1c}]+E[I_{2c}]} = \nonumber \\
    & =  \frac{1}{2\left(1-\cos\frac{2\pi}{N_\text{C}} \right)} / \left( \sum\limits_{j=2}^{N_1+1} \frac{1}{2\left(1-\cos\frac{2\pi}{N_\text{C}}j \right)} + \right. \nonumber \\
    & \left. + \sum\limits_{i \in C_1\cap C_2 \cap C_3} \frac{1}{\left(\sin^2\hat{O}_i + \frac{(h_\text{C}-h)^2}{(R_{\textrm{E}}+h_\text{C})^2} \cos\hat{O}_i\right)} \right),
\end{align}
\begin{equation}
    S_{2\text{C}} = \frac{P_{\text{Rxc}}}{E[I_{1c}]+E[I_{2c}]+kTB}.
\end{equation}

The main differences that modify the results slightly are:
\begin{itemize}
    \item First, the relative angular offset, $\Delta\beta(t)$, grows in the opposite direction. This causes the expected interference time profile to have even symmetry w.r.t. the expected interference time profile in the lower orbit.
    \item Second, the transmitter and receiver are further away from one another, which results in lower useful signal power at the receiver.
    \item Last, the interfering satellites located in the orbit of interest are further away from the receiver too. This causes the expected interference coming from the same orbit, $E[I_{1c}]$, to be lower with respect to the lower orbit.
\end{itemize}
The effects of these changes are further explored in Sec.~\ref{sec:results}.


\subsection{Shifted orbits}
\label{sec:shifted}
\begin{figure}[!h]
  \centering
  \vspace{-1mm}
  \includegraphics[width=0.38\textwidth]{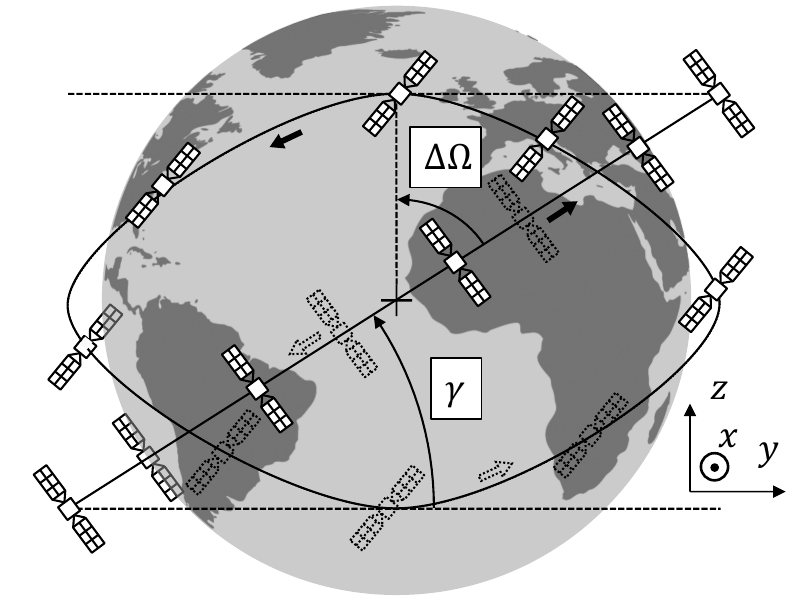}
  \vspace{-1mm}
  \caption{Modeling interference from a shifted orbit.}
  \label{fig:shifted_orbits_geometry}
  \vspace{-3mm}
\end{figure}

The third part of the model addresses the situation where the interference, $E[I_3]$, originates from an orbit with the same inclination and height as the orbit containing the link of interest, but a different $\Omega$. Figure~\ref{fig:shifted_orbits_geometry} depicts this scenario schematically, where $\Delta\Omega$ refers to the angular shift between the orbits. Typical Walker constellations combine multiple shifted orbits of this nature to provide continuous coverage between the latitude range $[-\gamma, \gamma]$~\cite{Leyva2022ngso}, as shown in Fig~\ref{fig:orb_2d}.
In this scenario, the need for a 3D coordinate system is clear, since satellites in the shifted orbits are not always within the same 2D plane. In particular, we use a Geocentric Equatorial Coordinate (GEC) system fixed with respect to the background stars. The $z$-axis is the axis of rotation of the Earth, the $x$-axis is the direction of the vernal equinox, and the $y$-axis follows a right-handed convention with the other two. We can identify the positions and pointing directions of each satellite using this coordinate system as 3-element vectors, which also allows us to calculate the angular separation between them.

We start by determining the location of each satellite within its own orbital plane for a given satellite $j$ out of $N_\text{S}$ in orbit 2 (orbit 1 contains the lawful transmitter and receiver). Concretely, we begin by computing its real anomaly (angular position within the orbital plane) as follows:
\vspace{-1mm}
\begin{equation}
  \phi_j = \Delta\beta + \frac{2\pi}{N_\text{S}}j,\ \forall j \in [0,N_\text{S}-1],
\end{equation}
\vspace{-1mm}
It is important to note that this time the relative angular offset $\Delta\beta$ is a constant parameter across time since both orbits have the same altitude and, consequently, the same orbital period. Typically, $\Delta\beta$ is treated as a constellation design parameter and it is fixed to a value that grants no collision between satellites at the orbit crossings. 

The position of the satellite in its own orbital plane, $\vec{r}_{0j}$, can then be obtained as:
\begin{equation}\label{eq:pos_orb_plane}
  \vec{r}_{0j} = (x_{0j}, y_{0j}, z_{0j}) = (R_{\textrm{E}}+h)(\cos \phi_j, \sin \phi_j, 0).
\end{equation}

To convert the position of the satellite relative to the orbital plane, $\vec{r}_{0j}$, to its coordinates in the GEC system, $\vec{r}_j$, the following conversion matrix is utilized:
\begin{equation}
  M_k = \begin{bmatrix}
    \cos\Omega_k & -\sin\Omega_k\cos \gamma & \sin\Omega_k\sin \gamma \\
    \sin\Omega_k & \cos\Omega_k\cos \gamma & -\cos\Omega_k\sin \gamma \\
    0 & \sin \gamma & \cos \gamma \\
  \end{bmatrix},
\end{equation}
where sub-index $k \in {1,2}$ is used to identify the orbit with the link of interest and the shifted orbit, respectively. $\gamma$ is the orbit's inclination, and $\Omega_k$ is the orbit's RAAN.
As $\vec{r}_j = M_k\vec{r}_{0j}$, we can then calculate the position of the $j$th satellite in the GEC system in orbit $k$. The same method is used to validate the interference conditions by finding the transmitter's and receiver's position vectors, $\vec{r}_{Tx}$ and $\vec{r}_{Rx}$, respectively.

In this scenario, the possible blockage by the Earth is analyzed with the rise and set function~\cite{Sharaf2012satellite} between the potential interfering satellite, $\vec{r}_j$, and the lawful receiver, $\vec{r}_{Rx}$, as follows:
\begin{align}
  \begin{split}
  R_j = & \langle \vec{r}_j, \vec{r}_{Rx}\rangle^2 - \|\vec{r}_j\|^2\|\vec{r}_{Rx}\|^2 \\
  & +(\|\vec{r}_j\|^2+\|\vec{r}_{Rx}\|^2)R_{\textrm{E}}^2 - 2R_{\textrm{E}}^2\langle \vec{r}_j, \vec{r}_{Rx}\rangle,
  \end{split}
\end{align}
thus establishing the set of visible satellites as:
\vspace{-2mm}
\begin{equation}
  C_1 = \{j \in [0,N_\text{S}-1]\ |\ R_j \leq 0 \}.
\end{equation}
\vspace{-6mm}

To find the sets $C_2$ and $C_3$, we calculate the distance vector between the receiver and the interfering satellite, and their respective pointing directions, as:
\vspace{-2mm}
\begin{align}
  \vec{r}_{Rx\rightarrow j} &= \vec{r}_{j}-\vec{r}_{Rx},\\
  \vec{r}_{Rx\rightarrow Tx}&= \vec{r}_{Tx}-\vec{r}_{Rx},\\
  \vec{r}_{j\rightarrow j-1} &= \vec{r}_{j-1}-\vec{r}_{j}.
\end{align}
\vspace{-2mm}
and angles $\psi_j$ and $\psi_j'$ are computed as:
\vspace{-1mm}
\begin{align}
  \psi_j &= \arccos \frac{\langle \vec{r}_{Rx\rightarrow j}, \vec{r}_{Rx\rightarrow Tx} \rangle}{\| \vec{r}_{Rx\rightarrow j} \| \cdot \| \vec{r}_{Rx\rightarrow Tx}\|},\\
  \psi_j' &= \arccos \frac{\langle \vec{r}_{j\rightarrow j-1}, \vec{r}_{j\rightarrow Rx} \rangle}{\| \vec{r}_{j\rightarrow j-1} \| \cdot \| \vec{r}_{j\rightarrow Rx}\|},
  \vspace{-1mm}
\end{align}
where $\vec{r}_{j\rightarrow Rx}=-\vec{r}_{Rx\rightarrow j}$.

The sets of satellites satisfying Conditions~2 and 3, $C_2$ and $C_3$, respectively, are then:
\vspace{-1mm}
\begin{align}
  C_2 &= \{j \in [0,N_\text{S}-1]\mod(N_\text{S})\ |\ |\psi_j| < \frac{\alpha}{2} \},\\
  C_3 &= \{j \in [0,N_\text{S}-1]\mod(N_\text{S})\ |\ |\psi_j'| < \frac{\alpha}{2} \}.
  \vspace{-5mm}
\end{align}

Finally, the total expected interference originating from the shifted orbit can be derived:
\vspace{-1mm}
\begin{equation}\label{eq:i3}
  E[I_{3}] = \sum_{j \in C_1\cap C_2 \cap C_3} \\ 
  \frac{\lambda^2 P_{Tx}}{4\pi^2 \left(1-\cos\frac{\alpha}{2}\right)^2 \|\vec{r}_{Rx\rightarrow j}\|^2},
\end{equation}
and taking into account the interference from the orbit of interest and the thermal noise, expressions for $\Gamma_3$ and $S_3$ can also be derived:
\vspace{-1mm}
\begin{align}\label{eq:sir3}
     & \Gamma_3 = \frac{P_{\text{Rx}}}{E[I_1]+E[I_3]} = \frac{1}{\left(1-\cos\frac{2\pi}{N}\right)} / \nonumber \\
     & / \left(\hspace{-1mm}\sum\limits_{i=2}^{N_1+1} \frac{1}{\left(1-\cos\frac{2\pi}{N}i \right)} +  2(R_{\textrm{E}}+h)^2\sum\limits_{j=0}^{N_j} \frac{1}{\|\vec{r}_{Rx\rightarrow j}\|^2}\hspace{-1mm} \right)
\end{align}
\vspace{-1mm}
\begin{equation}\label{eq:sinr3}
    S_3 = \frac{P_{\text{Rx}}}{E[I_{1}]+E[I_{3}]+kTB},
    \vspace{-1mm}
\end{equation}
where, again, expressions \eqref{eq:i1} and \eqref{eq:i3} are used for the calculation of $E[I_{1}]$ and $E[I_{3}]$ respectively.

\subsection{Co-planar Shifted Orbit}
\label{sec:shifted_coplanar}
A special extension of the analysis presented in the previous subsection is the model for the case where the shifted orbit is at a different altitude than the orbit with the target receiver, essentially consisting of a co-planar shifted orbit, as depicted in Fig.~\ref{fig:shifted_shifted_coplanar}. This is needed to model possible multi-altitude constellations designs, but more importantly, to properly model the interference among \emph{two different LEO constellations at neighbor altitudes (i.e., deployed by two different operators).}

Importantly, the approach presented in the previous subsection allows to do this by modifying \eqref{eq:pos_orb_plane} and including the interfering orbit altitude there, $h_{\textrm{S}}$, instead of the altitude of the orbit with the target receiver, $h$:
\vspace{-1mm}
\begin{equation}
  \vec{r}_{0j} = (x_{0j}, y_{0j}, z_{0j}) = (R_{\textrm{E}}+h_{\textrm{S}})(\cos \phi_j, \sin \phi_j, 0).
  \vspace{-1mm}
\end{equation}

It is important to note that because of the different altitudes, the relative angular offset between the target orbit and the shifted co-planar orbit will change over time obeying the dynamics derived in Appendix~\ref{app:rel_angular_offset}, as for co-planar orbits. The rest of the shifted orbit analysis stays the same.

\section{Numerical Results}\label{sec:results}
The numerical analysis of the mathematical models elaborated above is presented in this section. In addition to the results for each of the studied sub-scenarios (Sec.~\ref{sec:single_res} to \ref{sec:shifted_res}), in Sec.~\ref{sec:complete_res} we leverage the presented framework to analyze the numerical results when all three sub-deployments are combined as a single constellation.

Moreover, our numerical study focuses on two wireless communication bands. The first is millimeter wave (mmWave), which uses frequencies similar to those of 5G NR in terrestrial networks and is envisioned as a possible choice (by just adapting the existing hardware components design for space) for integrating Non-Terrestrial-Networks (NTNs) into the cellular infrastructure.

The second technology involves using the unlicensed sub-THz band to achieve ultra-broadband inter-satellite communication, which is becoming increasingly important as demand for higher data rates and the number of connected devices continues to grow. The selected modeling parameters for both bands are detailed in Table~\ref{tab:system_parameters}.

\subsection{Cross-check via Computer Simulations}
To ensure the accuracy of the delivered numerical results and the developed conclusions, the analytical models have been validated across computer simulations of the selected scenarios using an in-house developed satellite communications framework first utilized in~\cite{Alqaraghuli2021compact}. Our simulation platform, implemented in Python 3.8, carefully models the geometry, mobility, and orbital dynamics of moving satellites in defined constellations~\cite{Maral2009satellite}. It also captures the peculiarities of LEO orbital mmWave and THz wireless cross-links by following the models and assumptions from Sec.~\ref{sec:system_model}.
\begin{table}[!h]
    \footnotesize
    \centering
    \vspace{-1mm}
    \begin{tabular}{p{0.16\columnwidth}p{0.34\columnwidth}p{0.34\columnwidth}}
        \hline
        \textbf{Parameter}&\textbf{mmWave (K$_{\boldsymbol{a}}$ Band)
        }&\textbf{sub-THz} \\
        \hline
         $P_{Tx}$ & 60~dBm \cite{3gpp.38.821} & 27~dBm \cite{Siles2018new}\\
         $B$ & 400~MHz \cite{3gpp.38.821} & 10~GHz \cite{Sen2021versatile}\\
         $T$ & 100~K \cite{ITU372} & 100~K \cite{ITU372}\\
         $f_\text{C}$ & 38~GHz \cite{3gpp.38.821} & 130~GHz \cite{Siles2018new}\\
         \hline
    \end{tabular}
    \caption{Parameters utilized in the analysis for each frequency band.}
    \label{tab:system_parameters}
    \vspace{-2mm}
\end{table}

Our simulation platform is implemented in Python 3.8 and carefully models the geometry, mobility, and orbital dynamics of the moving satellites in defined constellations~\cite{Maral2009satellite}. The tool also captures the peculiarities of LEO orbital mmWave and THz wireless cross-links by following the models and assumptions from Sec.~\ref{sec:system_model}. Additionally, the tool can model satellite-to-ground and ground-to-satellite communications following the International Telecommunication Union (ITU) Recommendations ITU-R P.676-12~\cite{ITU676} and ITU-R P.835~\cite{ITU835}. It also allows for simulations involving non-trivial antenna radiation patterns and orientations, including the cone-plus-sphere (CPS) antenna pattern with multiple sidelobes pointing into different directions utilized in~\cite{Petrov2017interference} and several other studies. We utilize this more advanced CPS model for comparison purposes further.

The framework follows time-driven simulation approach. Specifically, at each timestep, the simulator propagates the orbit of all satellites in the constellation to their respective positions and computes the desired metrics of interest. Upon completion of the simulation, the results are stored for further post-processing and visualization.

\subsection{Single orbit}
\label{sec:single_res}
\begin{figure}[!h]
  \centering
  \subfigure[Signal-to-interference ratio]
  {
    \includegraphics[width=0.475\textwidth]{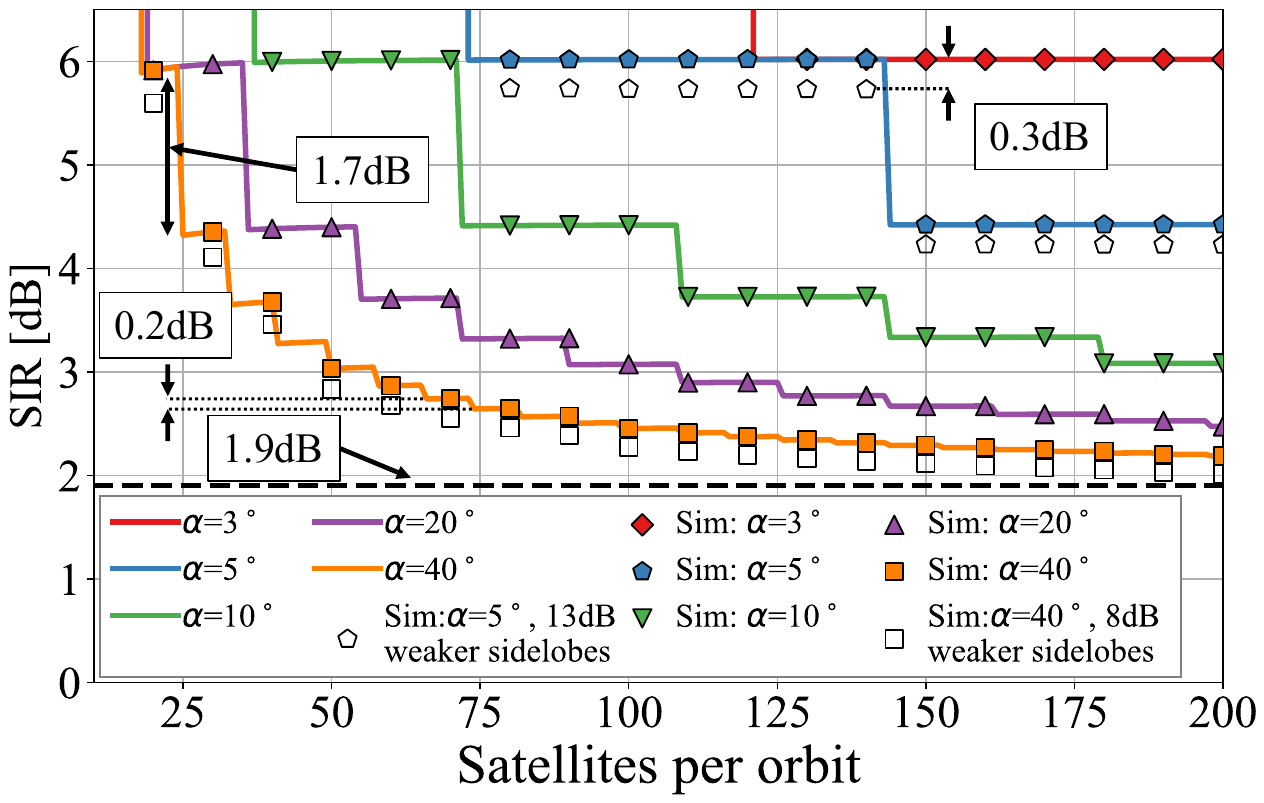}
    \label{fig:single_orbit_SIR}
     
  }
  \hfill
  \subfigure[Signal-to-interference-plus-noise ratio]
  {
    \includegraphics[width=0.475\textwidth]{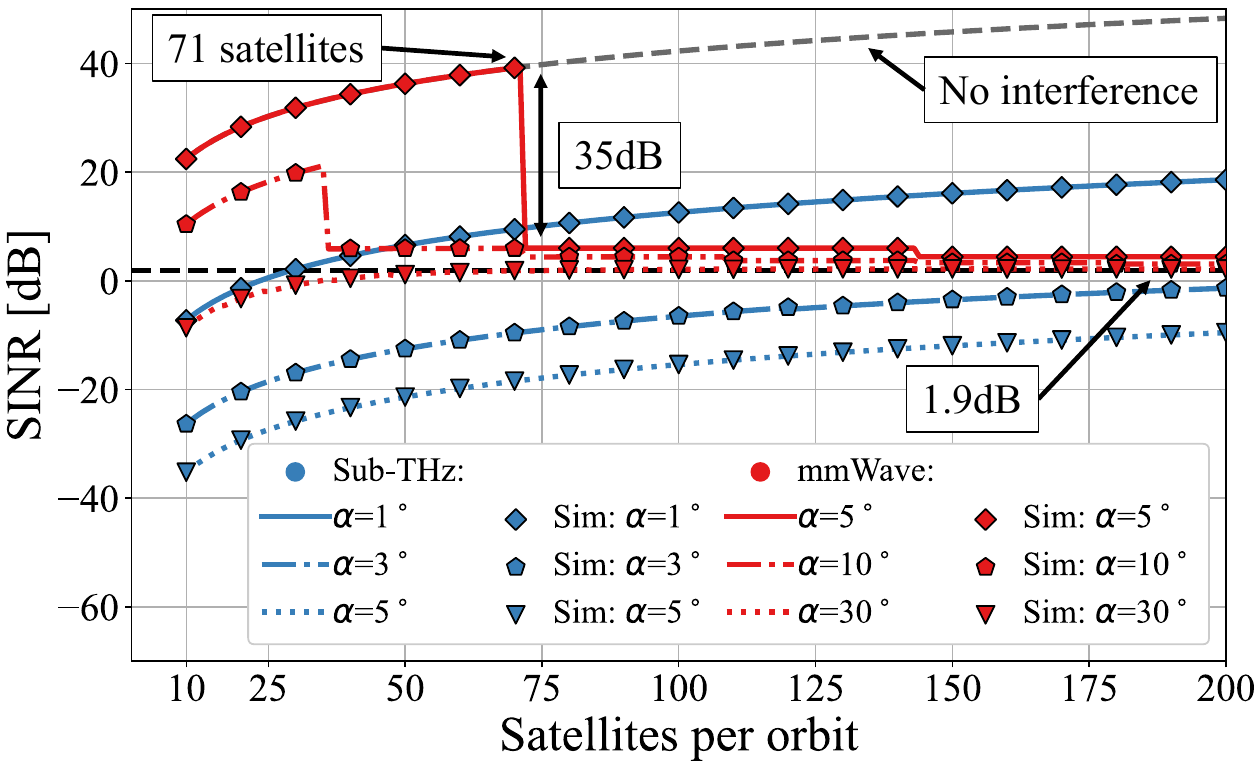}
    \label{fig:single_orbit_SINR}
     
  }
  \hfill
  \subfigure[Link capacity]
  {
    \includegraphics[width=0.475\textwidth]{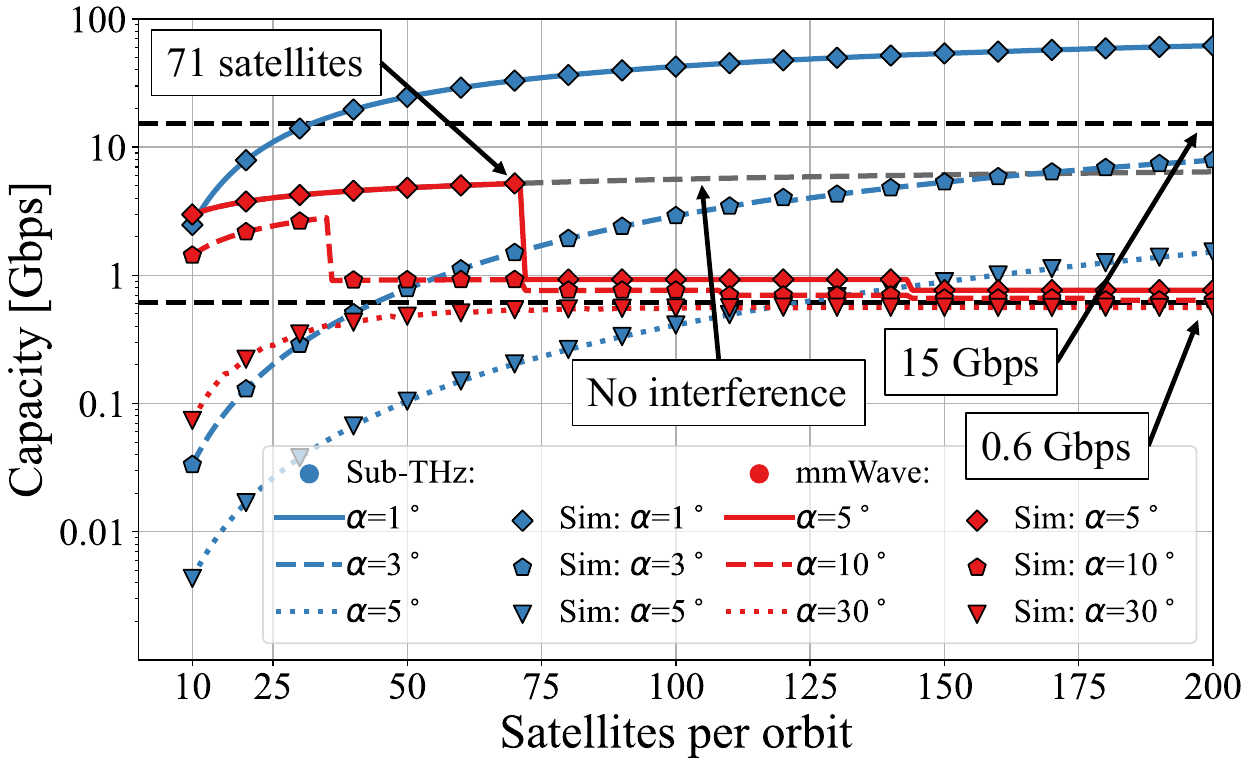}
    \label{fig:single_orbit_capacity}
     
  }
  \hfill
  \captionsetup{justification=centering}
  \vspace{-2mm}
  \caption{Single orbit constellation at $h=500$\,km.}
  \vspace{-3mm}
  \label{fig:single_orbit_SINR_capacity}
\end{figure}

We begin by examining the effects of the interference on the transmitted signal to isolate the signal impairments solely caused by interference in the system, and not thermal noise. Figure~\ref{fig:single_orbit_SIR} depicts the SIR for a single orbit setup, $\Gamma_1$, as a function of the number of satellites in orbit, $N$. Even though it may be unrealistic to consider so many satellites in an actual deployment, values of up to 200 satellites are analyzed to show the asymptotic behavior of the metric\footnote{Although the existing constellations are still not reaching such densities today, there is a clear trend among the main LEO satellite service providers in growing the number of satellites per orbit, eventually reaching these transition points if the process continues as of today. For instance, Starlink gen1 already has a set of polar orbital planes with 58 satellites per orbit~\cite{spaceXGen1}, while SpaceX has planned 120 satellites per orbital plane for Starlink gen2~\cite{spaceXGen2}. OneWeb initially planned 40~\cite{delportillo2019technical}, and recently updated that number to more than 70 satellites per orbit for the complete constellation~\cite{OneWeb, delportillo2019technical}. Project Kuiper System has planned 3,239 satellites operating in 98 orbital planes, resulting in at least 33 satellites per plane if they are evenly distributed~\cite{kuiper}. In China, the government and some companies are also planning to complete two massive constellations soon, code-named GW-2~\cite{gw2} and GW-A59~\cite{gwA59}, with more than 30 and 50 satellites per plane each, respectively.}. A dashed line is included at the theoretical bound derived in \eqref{eq:N_lim} (1.9\,dB) for reference. The orbit altitude is set to $500$~km, although as observed in \eqref{eq:sir1} and later confirmed with simulations, the orbit altitude, $h$, does not impact the average SIR, as the altitude-dependent parameters cancel each other.

The figure shows that, despite the reduction in transmitter-to-receiver distance, more satellites typically result in more interference and, as a result, lower SIR for a fixed beamwidth, $\alpha$. The trade-off between link distance and number of interfering satellites in LOS is reflected in the non-monotonic behavior of the curves. Although the SIR gets slightly better as the distance between Tx and Rx is reduced, i.e., more satellites are considered, Fig.~\ref{fig:single_orbit_SIR} shows precise drops in SIR from $k$ to $k+1$ satellites. These points occur when the number of interfering satellites rises as a result of growing the number of satellites. Noteworthy, the perceived drops are smaller as $N$ increases. For instance, the curve $\alpha=40\degree$ exhibits a drop in SIR from 24 to 25 satellites of $>$$1.5$\,dB, while the corresponding SIR decrease from 73 to 74 satellites is $<$$0.2$\,dB. As $N$ is increased, all the curves eventually converge to the 1.9\,dB bound denoted by the black dashed line.

We also note that satellite deployments are consistent with the intuitive relationship between SIR increasing with antenna directivity (confirmed in~\cite{HeathInterference} and~\cite{Petrov2017interference}, among other works). Nonetheless, even for $5\degree$ narrow-beam links, the SIR can be as low as $6$\,dB, posing a threat to the cross-link's performance and reliability if not adequately addressed. 

In addition to the simulated results, which closely match analytical predictions, we incorporate two further simulations adopting a more realistic antenna radiation pattern in Fig.~\ref{fig:single_orbit_SIR}. We modified the simulation tool to also model a pessimistic radiation pattern with uniform sidelobe power loss (CPS antenna radiation pattern used in~\cite{Petrov2017interference} and other works). In theory, this may lead to additional interference through sidelobes (thus lower SIR curves) than our model predicts.

However, after a careful study, it appears that any notable deviations (still fractions of dB) only appear when over $99\%$ of the signal power starts coming from the sidelobes for $\alpha=5^{\circ}$ and when over $80\%$ of the signal power comes from the sidelobes for $\alpha=40^{\circ}$. In practice, this translates to numerous sidelobes equally spread in all directions with only $13$\,dB and $8$\,dB weaker gains compared to the center of the main beam. This is a very pessimistic assumption reflecting a poor antenna design with very low spatial isolation. For instance, the widely adopted ITU model for interference studies~\cite{ITU_F699} utilizes the antenna model where sidelobes have $\approx$$30$\,dB lower gain compared to the main beam.

Hence, the non-negligible deviation in the curves behaviour in our study only appears with antenna models featuring $17$\,dB or more stronger sidelobes that recommended by the ITU, while also pointing these sidelobes into all the possible directions. Therefore, we observe that in the ovewhelming majority of practical setups, the presented analysis gives a good first-order estimations for SIR and SIR-dependent KPIs, while additional sidelobes (or other antenna radiation patterns) can still be considered if needed in follow up in-depth studies.

\begin{figure*}[!t]
  \centering
  \subfigure[SIR for a beamwidth $\alpha=10^\circ$]
  {
    \includegraphics[width=0.305\textwidth]{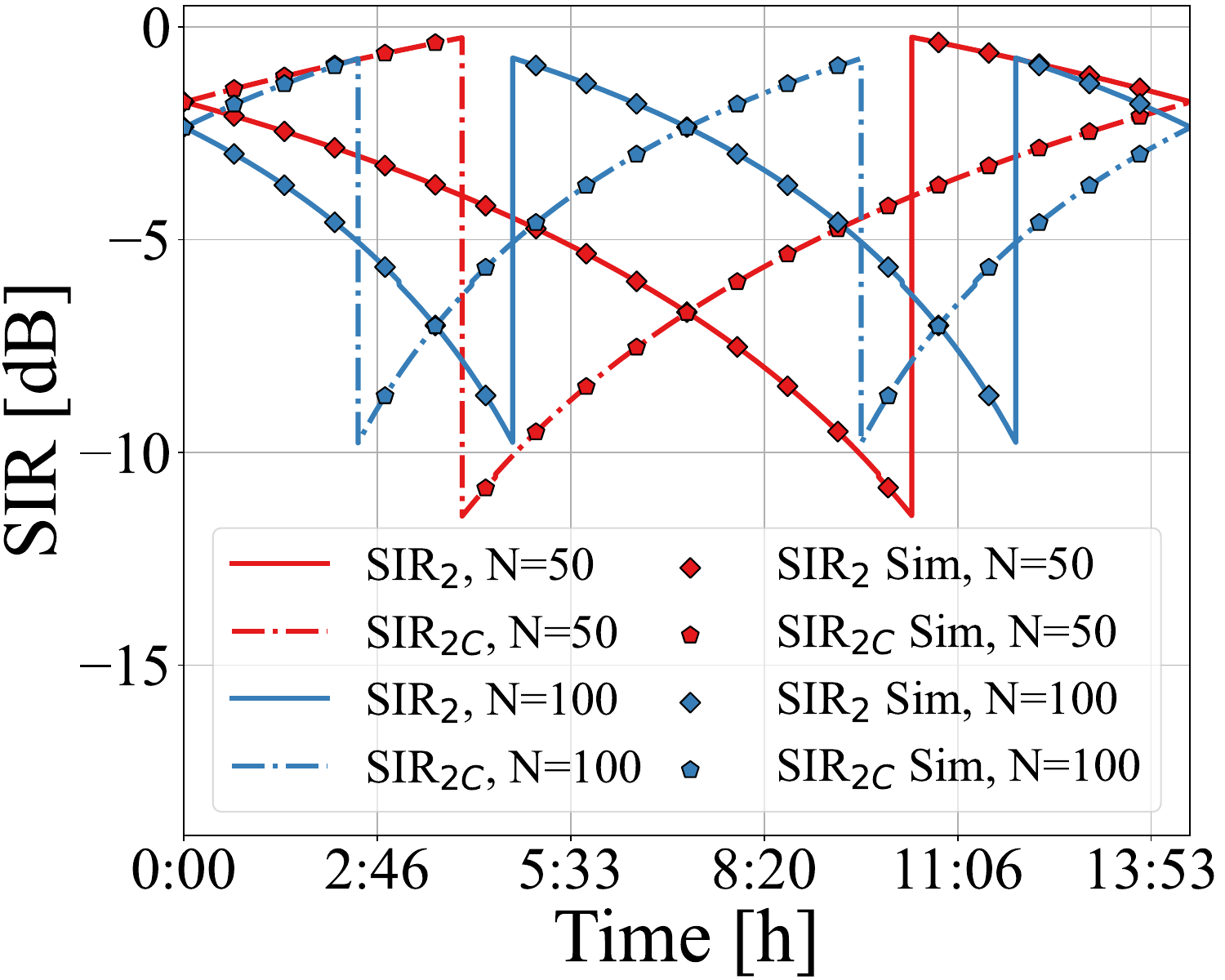}
    \label{fig:coplanar_orbits_SIR_alpha10}
     
  }
  \hfill
  \subfigure[SIR for a beamwidth $\alpha=30^\circ$]
  {
    \includegraphics[width=0.305\textwidth]{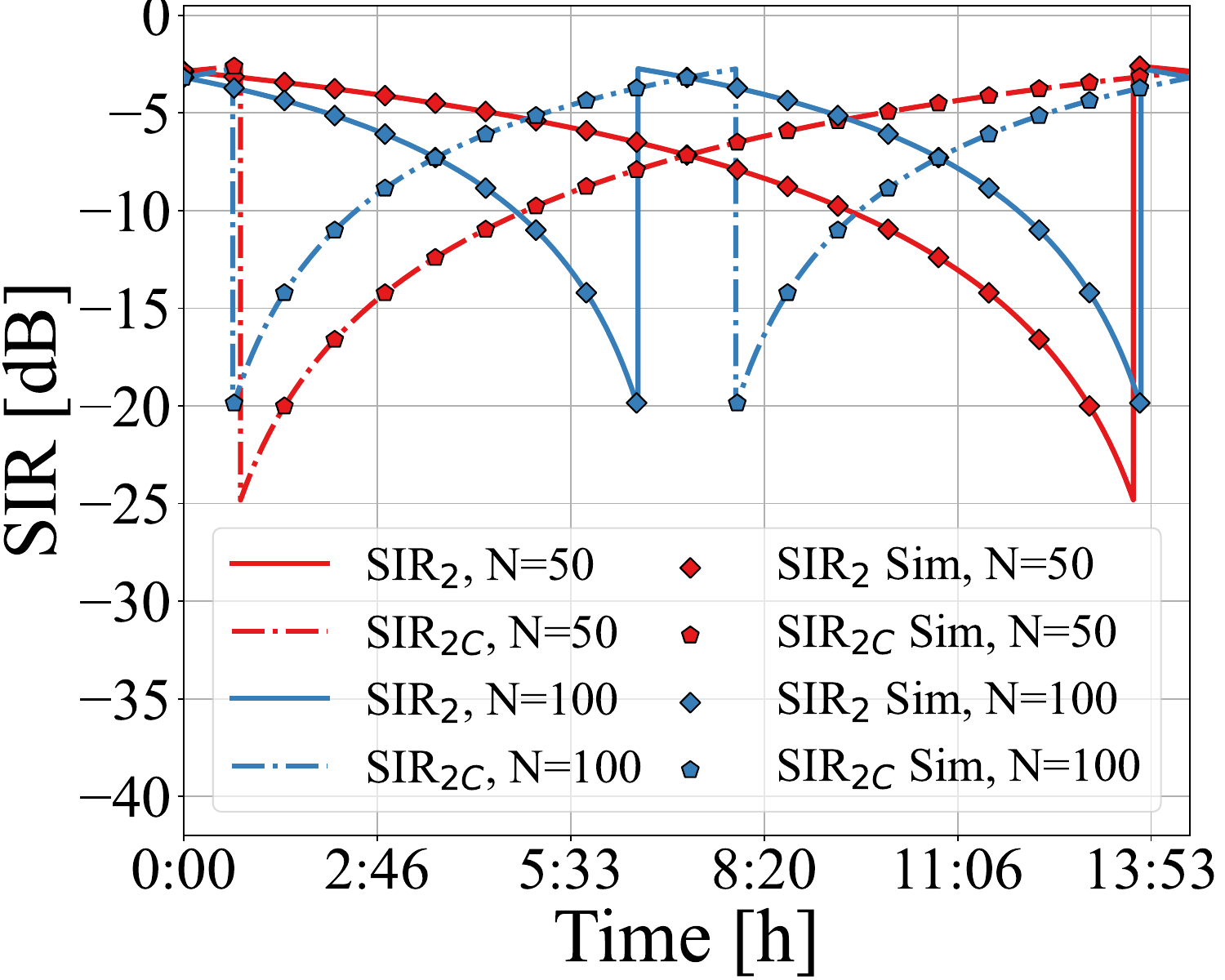}
    \label{fig:coplanar_orbits_SIR_alpha30}
  }
  \hfill
  \subfigure[Interference probability density]
  {
    \includegraphics[width=0.305\textwidth]{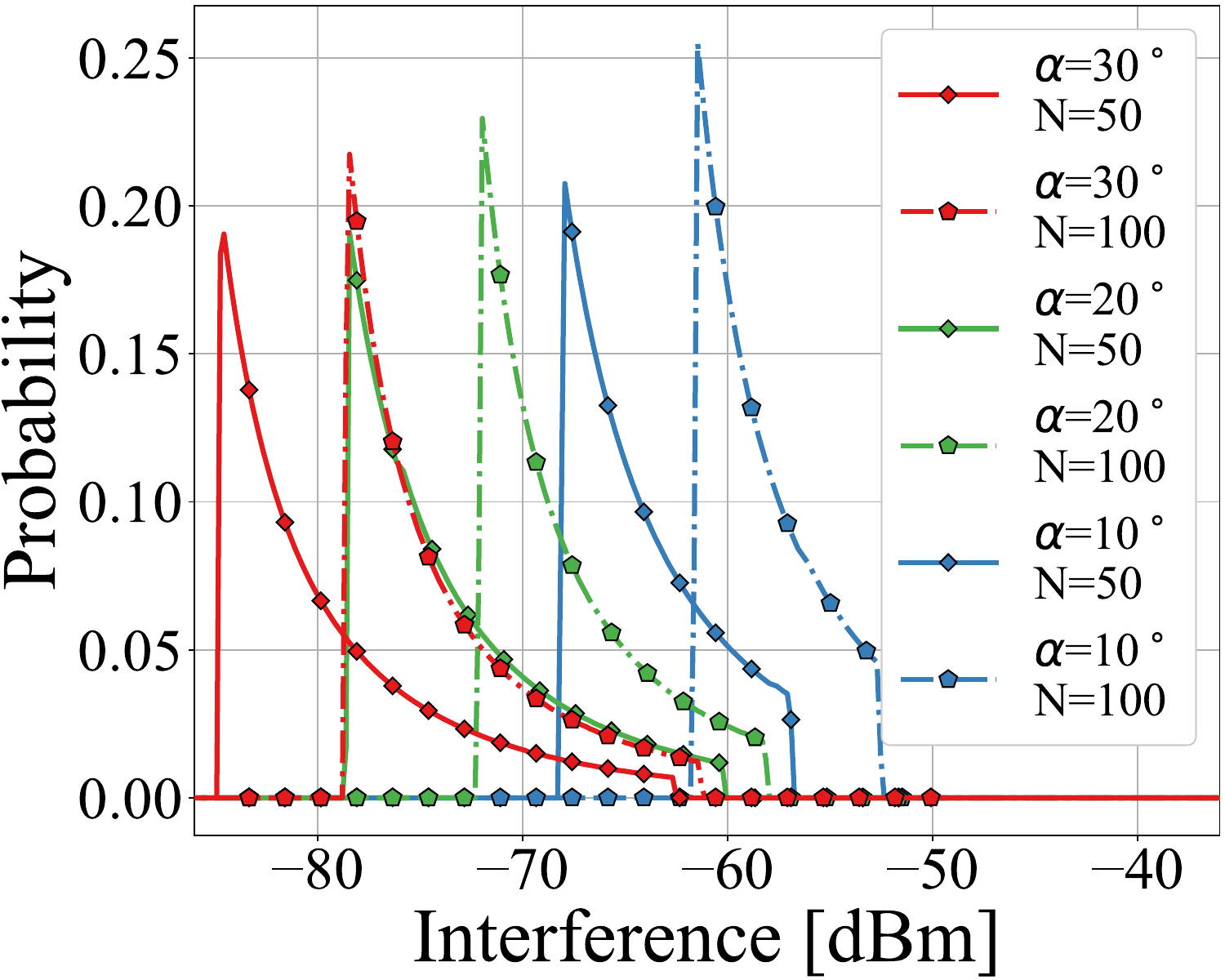}
    \label{fig:coplanar_orbits_I_pdf}
  }
  \captionsetup{justification=centering}
  \vspace{-2mm}
  \caption{Two co-planar orbits at $h=500$\,km and $h_\text{C}=510$\,km, respectively.}
  \label{fig:coplanar_orbits_SIR}
  \vspace{-6mm}
\end{figure*}

Further, in Fig.~\ref{fig:single_orbit_SINR} we incorporate the effect of the thermal noise into the analysis by studying the SINR in a single orbit deployment, $S_1$, and provide the corresponding channel capacity in Fig.~\ref{fig:single_orbit_capacity}. We first observe that, indeed, increasing the number of satellites $N$ effectively improves the link performance by reducing the distance between the transmitter and receiver only when there is no interference. For instance, the curves corresponding to THz depict a smooth increasing behavior because they are either not affected by interference  ($\alpha=1^\circ$), or because the effect of noise is dominant over the interference ($\alpha=3^\circ, 5^\circ$), for any considered value of $N$.

In general, due to the higher output power, we observe a better SINR of mmWave over THz, even though broader beams are considered. For instance, at the lower range of values of $N$, the performance for a link at mmWave with $\alpha=5^\circ$ is unmatched. However, the performance of that link is clearly compromised by interference at higher values of $N$. We observe a drop of more than 35\,dB in SINR at $N=72$ satellites, which is caused by the first interfering satellite appearing in LOS of the receiver. Therefore, in this setup ($h=500$\,km and mmWave cross-links with $\alpha=5^\circ$), $N^*=71$ is the optimal number of satellites to be deployed to maximize the link performance, as obtained in \eqref{eq:N_optimal}. Furthermore, as $N$ increases, we again observe that the SINR is eventually bounded by the SIR limit of 1.9\,dB, as obtained in \eqref{eq:N_lim}.

This bound is translated to an achievable capacity of around 600\,Mbps at mmWave and 15\,Gbps at THz, both indicated with black dashed lines respectively. From Fig.~\ref{fig:single_orbit_capacity} we deduce that the mmWave links fall in a bandwidth-limited regime, despite having a better SINR, which means that the capacity is also constrained by bandwidth. For example, even if no interference was present in the mmWave link with $\alpha=5^\circ$ (black dashed line), the channel capacity would still be lower than at THz despite having a considerably larger SINR. On the other hand, the channel capacity at THz rapidly increases when abandoning the power-limited regime (low $N$), achieving channel capacities of up to 80\,Gbps.

\subsection{Co-planar orbits}
\label{sec:coplanar_res}
Extending the single orbit scenario, a two co-planar orbits setup is analyzed next.

\subsubsection{Time-dependent interference}
In this setup, the SIR, as well as the other metrics of interest, change over time as a result of the satellite's relative motion and orientation in different orbits. In Fig.~\ref{fig:coplanar_orbits_SIR}, two co-planar orbits with $N=N_\text{C}$ and altitudes of $500$\,km and $510$\,km (thus, with only $10$\,km separation distance) are used to demonstrate this effect for two different antenna beamwidths. 

First, we observe that despite the relatively close distance between the orbits, the average SIR, $\Gamma_2$ and $\Gamma_{2\text{C}}$, exhibit periodic variations between $0$\,dB and $-25$\,dB, with sharp jumps when a new interfering satellite appears in LOS. We notice that the period changes depending on the number of satellites in the orbits, becoming much shorter as the number of satellites increases. This is due to the fact that a higher number of satellites per orbit translates to a smaller angular separation between them, resulting in a shorter relative rotation required to reach a redundant satellite position. The same effect is present in the co-planar orbit, which also exhibits even symmetry with its lower altitude counterpart. Notably, in addition to the moments where $\Gamma_2\approx \Gamma_{2\text{C}}$, there are instants at which $\Gamma_2<<\Gamma_{2\text{C}}$, which could be used as transmission windows by an intelligent routing algorithm. Furthermore, we notice that the beamwidth is crucial once more because curves for narrower beamwidths consistently outperform the broader beam curves.

If the Keplerian elements of the constellation are known, as is the case in Fig.~\ref{fig:coplanar_orbits_SIR}, the time dependence of the interference in this configuration can be well captured by the orbit dynamics. The two orbits could, however, belong to two separate satellite operators who have little to no knowledge of each other's satellites' whereabouts at any given time. In this case, it is appropriate to consider the expected interference at each operator's constellation as a uniformly distributed variable across time which has an associated probability density function (PDF). Fig.~\ref{fig:coplanar_orbits_I_pdf} depicts such PDFs at one of the orbits for different beamwidths $\alpha$ and number of satellites $N$, revealing the trade-off between $N$ and $\alpha$ for this setup. 

Here, the effect of directivity is reflected as a higher interference at narrower beamwidths, although the resulting SIR is compensated by the corresponding increase in signal strength, as seen in Fig.~\ref{fig:coplanar_orbits_SIR}. Interestingly enough, the shape of the different PDFs is similar, with low interference values showing the largest probability, leading to the conclusion that transmission windows are more likely than outage periods. Remarkably, the shape of the PDFs is also affected by the constellation parameters, with the largest variance at a low number of satellites and broader beams ($\alpha=30$ and $N=50$). The small distance between orbits, however, has an important impact on these findings.

\subsubsection{Time-averaged interference}
\begin{figure}[ht!]
\vspace{-2mm}
  \centering
  \subfigure[Signal-to-interference-plus-noise ratio]
  {
    \includegraphics[width=0.475\textwidth]{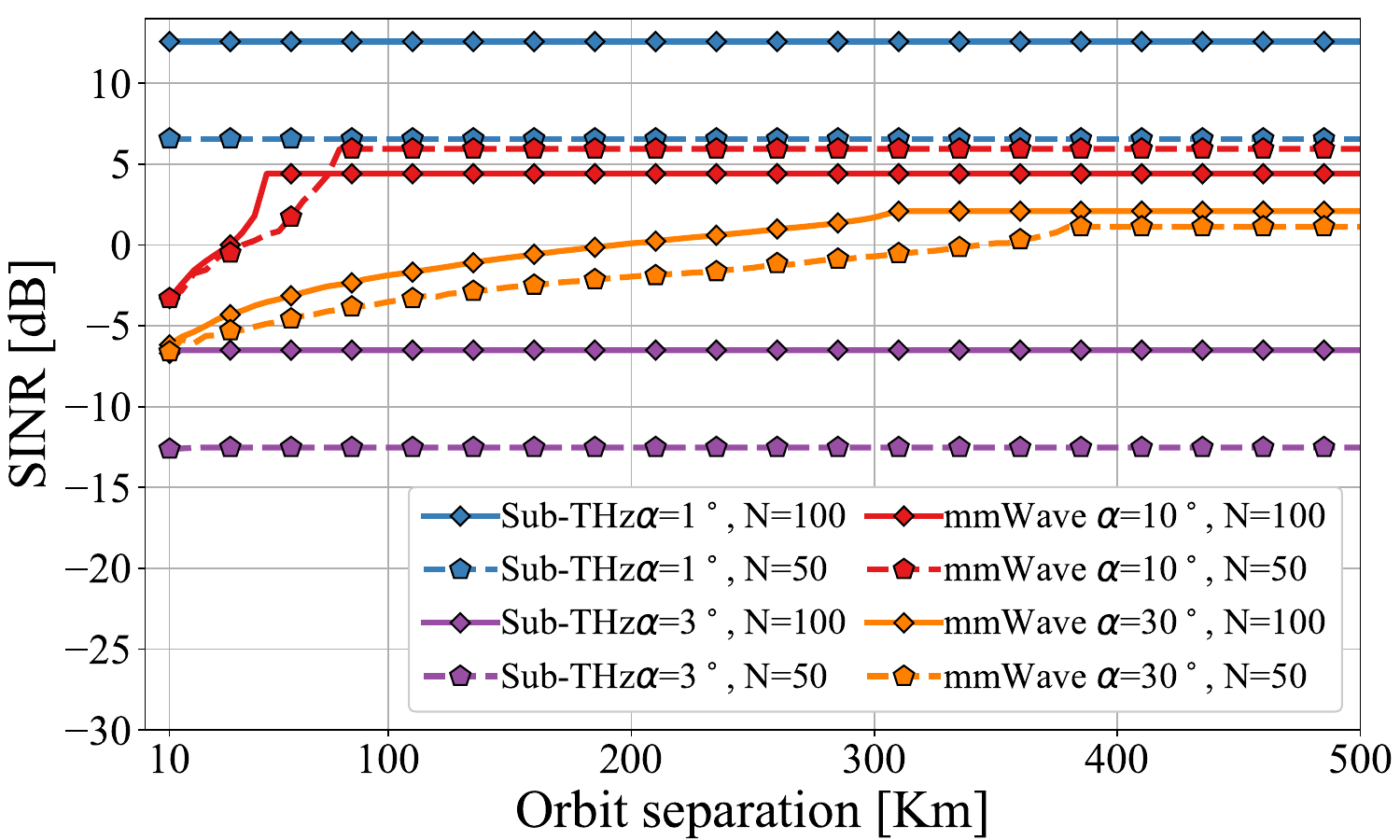}
    \label{fig:coplanar_orbits_SINR}
     
  }
  \hfill
  \subfigure[Channel capacity]
  {
    \includegraphics[width=0.475\textwidth]{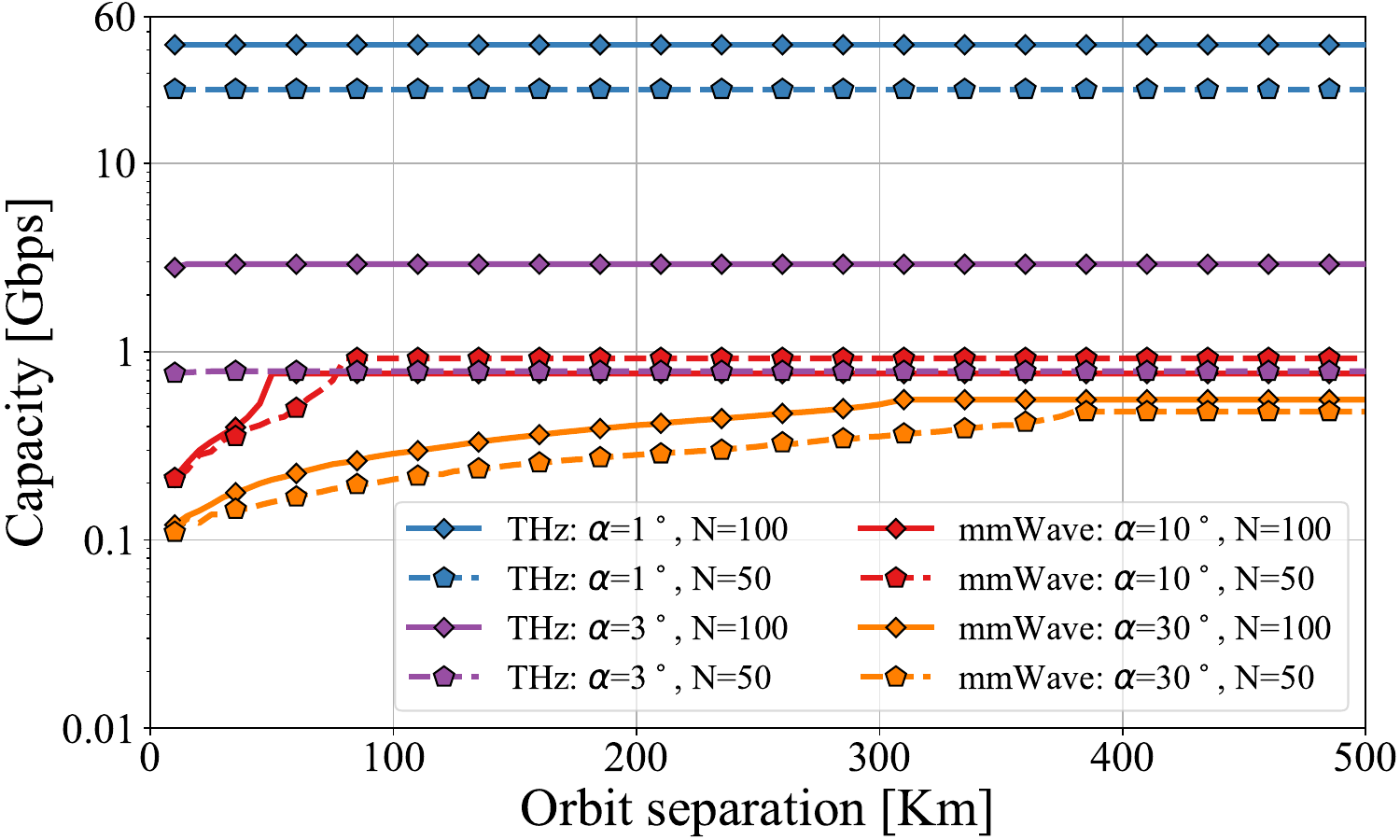}
    \label{fig:coplanar_orbits_capacity}
  }
  \hfill
  \captionsetup{justification=centering}
  \vspace{-2mm}
  \caption{Time-averaged results in a co-planar orbits setup using mmWave and THz cross-links.}
  \vspace{-2mm}
  \label{fig:coplanar_orbits_SINR_capacity}
\end{figure}

To explore the impact of the separation distance between orbits, in Fig.~\ref{fig:coplanar_orbits_SINR_capacity} we average the results across the period of observation, focusing on the final metrics of interest, i.e. SINR and channel capacity. Once more, we see that for THz links, communication in either of the two orbits is unaffected by interference due to the narrower beams necessary for closing the link budget. On the other hand, the mmWave links suffer degraded performance at low orbit separations due to interference from the co-planar orbit. Moreover, a crossing of the pair of curves with the same beamwidth at mmWave can be observed depending on the constellation parameters. For $\alpha=10^\circ$, the SINR for $N=100$ rapidly increases as the orbit separation is higher, eventually only capturing inner-orbit interference and, therefore, achieving a constant maximum just below $5$\,dB. On the other hand, the SINR for $N=50$ has a slower growth but reaches a larger SINR maximum when the link of interest is only affected by satellites in the same orbit. 

Interestingly, this crossing is not present in the mmWave curves for $\alpha = 30$. This effect is a consequence of the single orbit results and can be explained through Fig.~\ref{fig:single_orbit_SINR}. In the figure, the mmWave curve for $\alpha=10^\circ$ curves over the horizontal asymptote and eventually converges to it from above, decreasing as $N$ grows after the local maximum. Contrarily, a mmWave curve for a wider beamwidth, i.e. $\alpha=40^\circ$, remains below the asymptote due to a constrained SNR. This results in a monotonically increasing curve as $N$ grows until it also converges to the asymptote, but from below.

\begin{figure}[!h]
    \centering
    \includegraphics[width=0.475\textwidth]{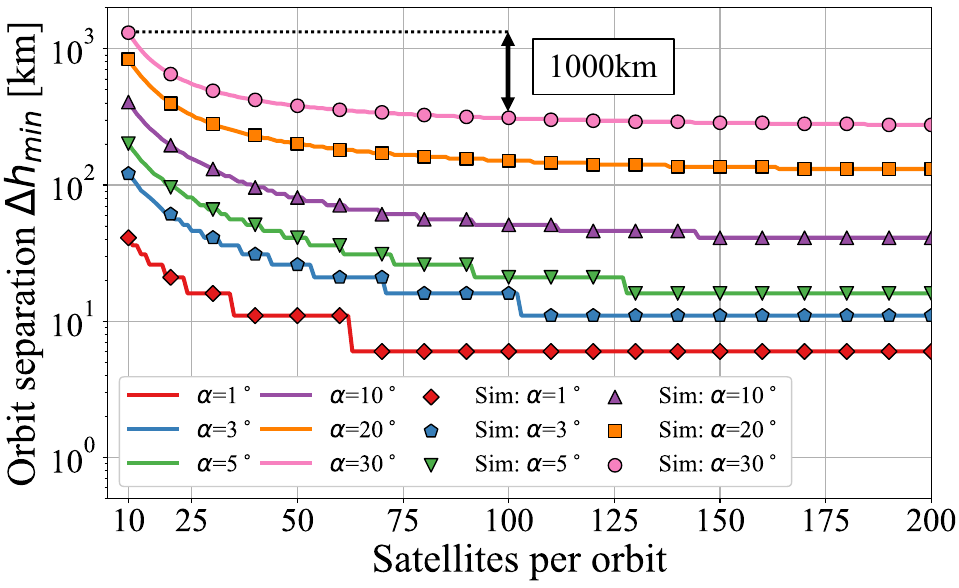}
    \caption{Required upper orbit separation to avoid interference with a co-planar orbit at $h=500$\,km}
    \label{fig:coplanar_orbits_min_h}
    \vspace{-3mm}
\end{figure}

This trend is better captured in Fig.~\ref{fig:coplanar_orbits_min_h}, which explores the minimum required orbit separation, $\Delta h_{min} = h_{\text{C}}^*-h$, for complete interference isolation between the two orbits, obtained through solving \eqref{eq:h_c_star}. Moving from left to right in the figure, we note that broad beams necessitate a large orbit separation for interference isolation. For example, for $\alpha=30^{\circ}$, an altitude separation of over $\Delta h = 1000$\,km is required for the lowest satellite count, potentially extending beyond the standard LEO range ($h_{\text{C}}<2000~\text{km}$). With narrower beams, such as $\alpha=1^{\circ}$, the required separation distance significantly decreases, often to less than $10$\,km, thus reducing the need for interference mitigation strategies in such setups and allowing for a full reuse of frequency channels among those two co-planar constellations.

Interestingly, we found that the minimum orbit separation for avoiding interference decreases as the number of satellites increases. For a beamwidth of $\alpha=3^{\circ}$, the required separation drops by nearly 1000 km when the number of satellites in orbit increases from 10 to over 100. This reduction is due to the change in satellite pointing directions towards the orbit's tangent, creating a larger angular separation from satellites in the adjacent orbit. In summary, while adding more satellites to a constellation may increase inner orbit interference, it can also enhance protection against interference from co-planar orbits of different operators.

Finally, because of the abundant bandwidth and sufficient directivity to close the link budget and avoid interference with significantly less power than at mmWave, the channel capacity for THz lines is once more unmatched, as detailed in Fig.\ref{fig:coplanar_orbits_capacity}.

\subsection{Shifted orbits}\label{sec:shifted_res}
We proceed by numerically modeling the interference from a shifted orbit/orbits.

\begin{figure*}[!h]
  \centering
  \subfigure[Signal-to-interference ratio]
  {
    \includegraphics[width=0.305\textwidth]{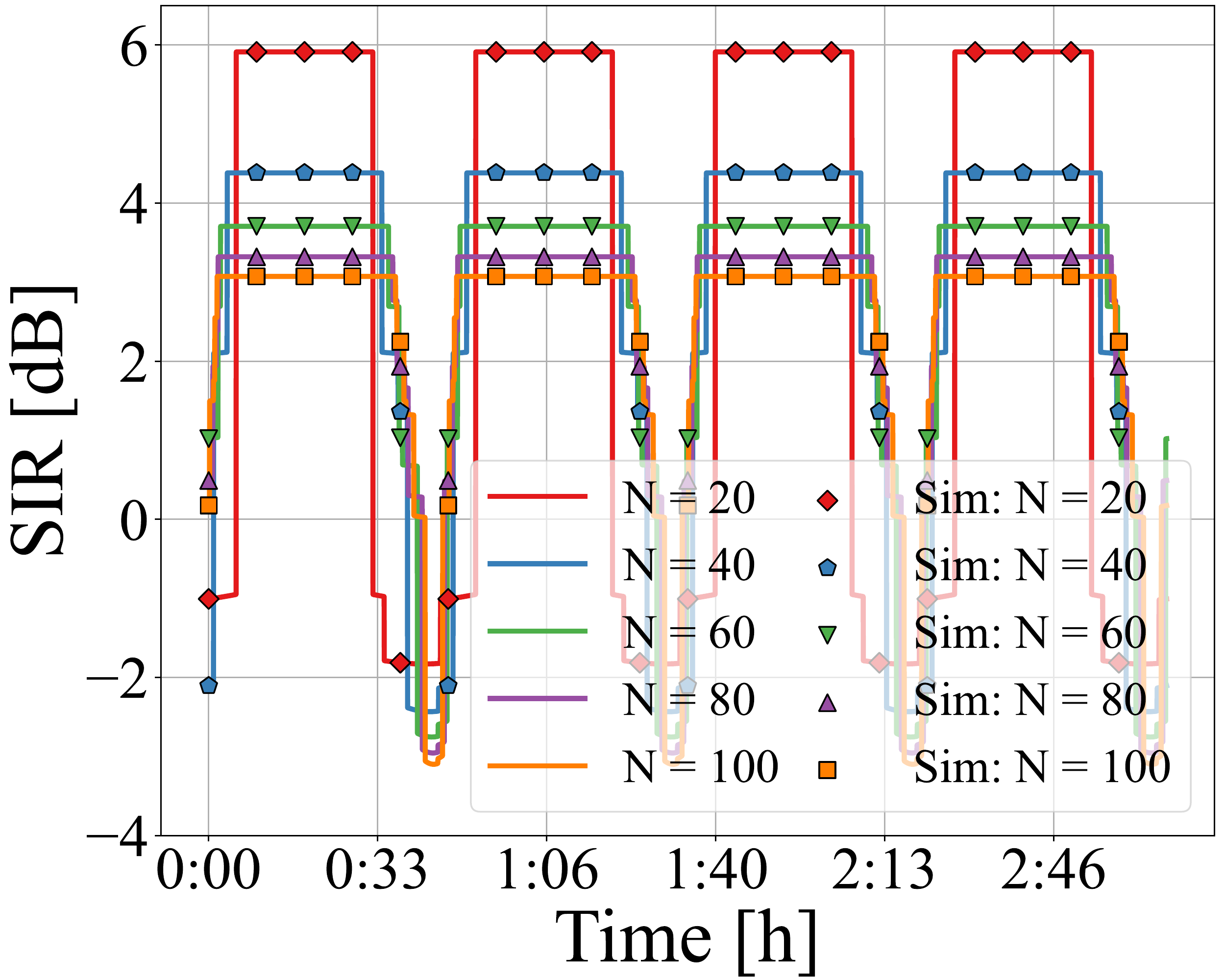}
    \label{fig:shifted_orbits_SIR}
     
  }
  \hfill
  \subfigure[Signal-to-interference-plus-noise ratio]
  {
    \includegraphics[width=0.305\textwidth]{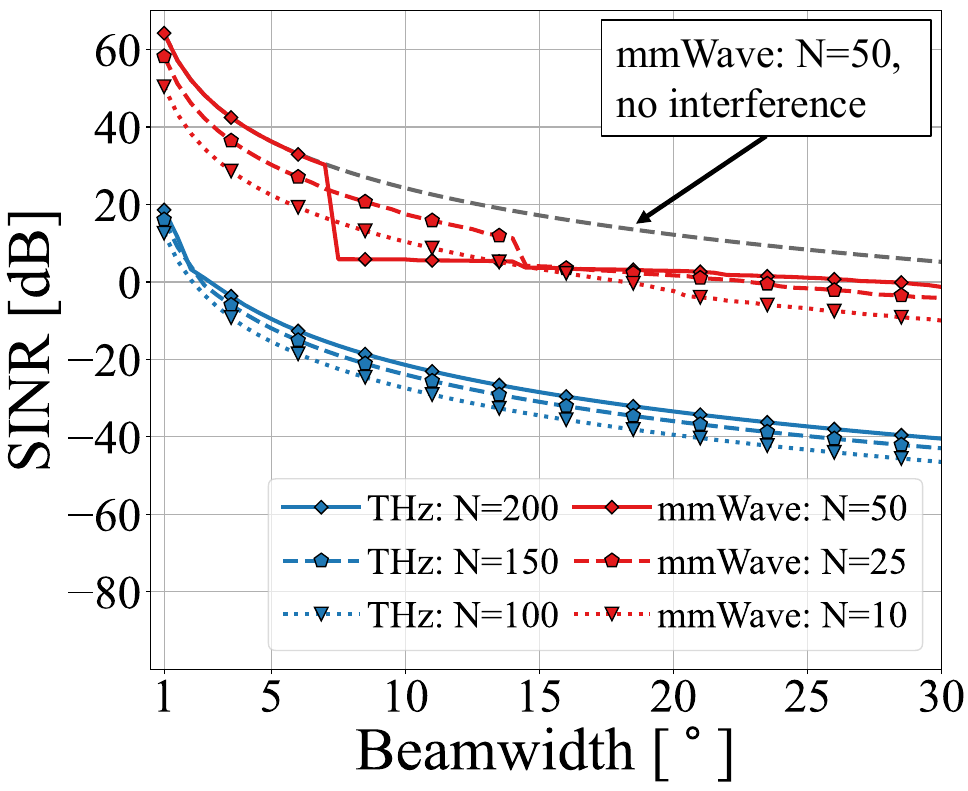}
    \label{fig:shifted_orbits_SINR}
  }
  \hfill
  \subfigure[Link capacity]
  {
    \includegraphics[width=0.305\textwidth]{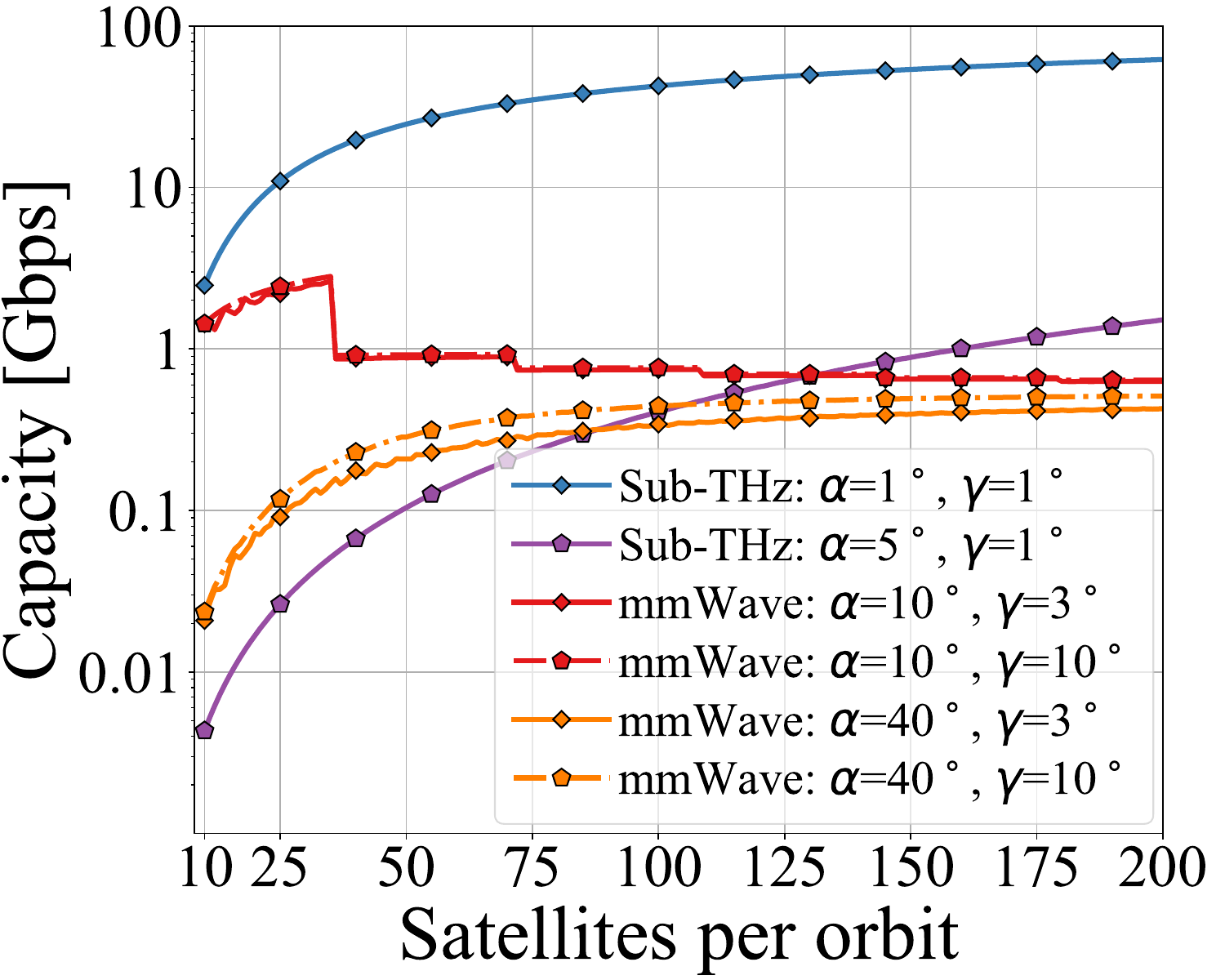}
    \label{fig:shifted_orbits_capacity}
  }
  \hfill
  \captionsetup{justification=centering}
  \vspace{-2mm}
  \caption{Two shifted orbits at $h=500$\,km and a RAAN shift $\Delta\Omega=90^{\circ}$.}
  \vspace{-6mm}
  \label{fig:shifted_orbits}
\end{figure*}

\subsubsection{Time-dependent interference}
As in the co-planar setup, the relative motion between satellites at different orbits causes time variations in the captured interference. Of special importance in this scenario are the moments where the two orbits cross, which result in performance degradation, as shown in Fig.~\ref{fig:shifted_orbits_SIR}. The figure depicts the SIR for a setup with a $3^{\circ}$ inclination, $N$$=$$N_\text{S}$, and $\alpha=30^{\circ}$, and a time span equal to two orbital periods. Similar to the single orbit scenario, the SIR in this case is frequently only impacted by satellites in the same orbit. However, we observe brief outage periods during the time periods associated with the orbital crossings, and throughout which transmission should be avoided. We again observe periodicity in the curves, with two distinct SIR decreases per cycle (four in the figure, since the interference cycle matches one orbital period).

Interestingly, the time span of the SIR drops is narrower as the number of satellites in orbit increases. This reflects that despite having more satellites in LOS, the shorter distance between transmitter and receiver allows for a longer sustained SIR period and a slightly more gradual drop in SIR when getting closer to the satellites in the opposite orbit.

\subsubsection{Time-averaged interference}
To study the impact of the constellation parameters on the metrics of interest, the results have been averaged over time once more. Fig.~\ref{fig:shifted_orbits_SINR} depicts the SINR, $S_3$, as a function of the beamwidth $\alpha$ for the two considered bands. As expected, the SINR at mmWave is consistently higher than at THz because of the larger available power, although as seen before, larger SINR does not translate into larger capacity. Moreover, despite the lower number of satellites considered, the mmWave links get rapidly affected by interference. A grey line corresponding to the SNR for $N=50$ is added for reference to reflect the impact of the SINR drop around $\alpha=7^\circ$. Even though the resulting SINR is still acceptable, the limitation in the bandwidth of the mmWave links results in a considerably lower channel capacity. In contrast, the THZ links exhibit considerably lower link performance, outlining again the necessity of directional links to close the link budget with decent SINR values. The smoothness of the THz curves is due again to either the lack of interference (SINR$=$SNR) or the huge impact of noise masking that of the present interference (SINR$\approx$SNR).

The effect of increasing the number of satellites in orbit is also explored in Fig.~\ref{fig:shifted_orbits_capacity} for different beamwidths, $\alpha$, and orbit inclination, $\gamma$. Compared with Fig.~\ref{fig:single_orbit_capacity} from the single orbit scenario, the curves for the THz band are identical since they are unaffected by interference, even at the lowest inclination angle possible. On the contrary, the mmWave links suffer slight performance degradation due to the presence of the shifted orbit, especially with broad beamwidths, e.g. with $\alpha=40^\circ$. Of special interest is the ripple effect that starts to appear because of the shifted orbit presence. The origin of this effect is found in the drops in SIR shown in Fig.~\ref{fig:shifted_orbits_SIR}. We observe that small increments in the number of satellites directly affect the shape and depths of the drops, causing the "ripple" effect in the time-averaged results. Overall, however, the effect of a shifted orbit in the cross-link interference can be considered minimal compared with the interference from the same orbit, which confirms the usefulness of shifted orbits as the building block for Walker constellations shells.

\subsection{Combined realistic deployment: Lessons learned}\label{sec:complete_res}
\begin{figure*}[!t]
  \centering
   \includegraphics[width=0.98\textwidth]{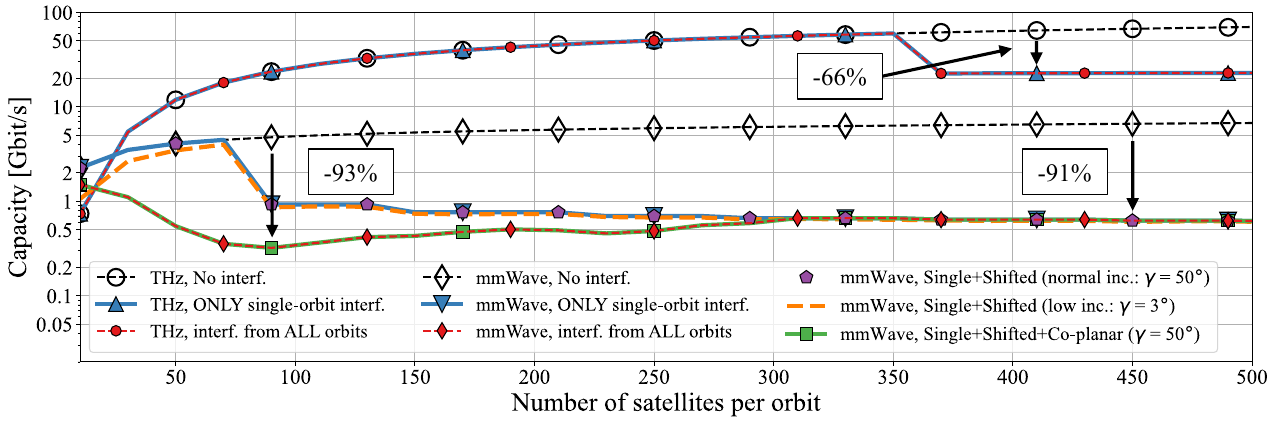}
  \vspace{-1mm}
  \caption{Full two-constellations setup. Modeling two realistic Walker constellations at $h=500$~km and $h_\text{S}=510$~km, respectively, each of $10$ evenly spaced shifted orbits with a $50^\circ$~inclination. Exploring the sub-THz radio with $\alpha=1\degree$ and the mmWave radio with $\alpha = 5\degree$ (the rest of parameters are as per Table~\ref{tab:system_parameters}).}
  \label{fig:complete_constellation_capacity}
  \vspace{-5mm}
\end{figure*}

As a final part of the numerical study, the interference in a complete constellation deployment is analyzed in Fig.~\ref{fig:complete_constellation_capacity} by combining all four major sources of interference: from the same orbit (``single orbit''), from a co-planar orbit (``co-planar''), from shifted orbits (``shifted''), and from shifted co-planar orbits (``shifted co-planar''). While each of the subsections below aims to provide an in-depth study of the impact and dynamics of particular interference sources (e.g., from the same orbit or a co-planar orbit), Fig.~\ref{fig:complete_constellation_capacity} presents all these sources together in a realistic large-scale deployment, also comparing the impact of different sources against each other at a high level.

A schematic view of a realistic constellation is given in Figure~\ref{fig:orb_2d}, where only two orbital planes are considered for illustrative purposes. For this study, a practical-size constellation is modeled consisting of $10$ evenly spaced orbital planes at an inclination of $50^\circ$ and an altitude of $500$~km. Each of the orbits features from $10$ to $500$ small satellites, making it from $100$ (low density) to $5000$ satellites (extreme case) per constellation. Additionally, a similar constellation is deployed at a $10$~km higher orbit ($510$~km altitude), hence modeling possible co-existence between two full satellite constellations (e.g., by two different operators). We model both sub-THz and mmWave radio setups as per Table~\ref{tab:system_parameters} and focus on the achievable capacity for a cross-link as a function of the number of satellites per orbit.

Starting from sub-THz results, we observe that the use of narrow THz beams ($\alpha=1^{\circ}$) allows to keep the setup completely interference-free all the way until $350$ satellites per orbit. Then, the interference from the same orbit appears and causes the deviation between the theoretical results for interference-free regime (capacity based on SNR) from a more realistic setup, where the capacity limit is determined by the SINR. Importantly. the only non-negligible source of interference from $350$ to $500$ satellites per orbit with $1^{\circ}$-wide sub-THz beams is from the same orbit (single orbit setup). Following the scenario geometry, other sources of interference (shifted, co-planar, and shifted co-planar) appear only with very high number of satellites per orbit (over $1000$). \emph{Hence, when modeling narrow-beam sub-THz and THz cross-links, it is important to account for the interference from the same orbit, while all other sources of cross-link interference may be ignored for first-order analysis in many cases.}

Continuing further with mmWave radio setup and $\alpha=5^{\circ}$ beamwidth, a similar relation between ``No interference'' and ``Single-orbit'' results is noted, yet curve deviations occur at a lower density (around $70$ satellites per orbit). \emph{Importantly, in mmWave cross-links, interference isn't limited to same-orbit satellites. Particularly, the interference from shifted orbits is of secondary importance in high inclination scenarios, but must be taken into account for low-inclination setups (e.g., with $\gamma = 3\degree$). Moreover, with only $10$~km separation between two orbits, the co-planar interference becomes a major issue preventing the capacity of the mmWave cross-link from growing over $0.8$ Gbit/s, which is insufficient for prospective high-rate-centric satellite networks. Hence, the interference from a co-planar orbit must be accounted for in designing efficient policies for constellation deployments, spectrum allocation, and static/dynamic spectrum sharing among different satellite constellations or different satellite network operators.}



\section{Conclusions and discussion}
\label{sec:conclusions}

As a crucial component of 5G-Advanced and 6G networks, broadband directional cross-links over mmWave and future THz bands are key technological enablers for high-rate LEO satellite communication services. This paper introduces an analytical framework for modeling cross-link interference in such systems. The framework is utilized to model three tractable scenarios that are the key building blocks of future LEO constellations, and their combination in realistic multi-constellation deployments. All results have been validated by extensive simulation using our in-house developed simulator.

For the studied scenarios, we have characterized the impact of the antenna directivity and the constellation parameters on the aggregate interference, SIR, SINR, and channel capacity in massive satellite deployments. Key findings include:
\begin{enumerate}
\vspace{-2mm}
\item Cross-link interference significantly affects KPIs in  realistic deployments and therefore cannot be neglected.
\item The primary source of interference arises from satellites within the same orbit, which will require special consideration from satellite operators planning to deploy extensive LEO satellite constellations.
\item The relative motion and orientation of satellites in different orbits cause the interference to vary over time, offering opportunities for intelligent routing algorithms to exploit periods of temporarily reduced interference.
\item There is a trade-off between utilizing mmWave and THz bands for ultra-broadband cross-links. MmWave systems feature higher output power, which eliminates the need for extremely narrow beams ($<5^\circ$) and the associated pointing challenges. MmWave band is particularly better for a smaller number of satellites per orbit, where cross-links are hundreds of kilometers long. In contrast, when range is not an issue (hundreds of satellites per orbit), narrower sub-THz beams offer larger bandwidth and are much less vulnerable to interference from other cross-links, thus resulting in superior channel capacity.
\end{enumerate}

Besides the key findings and observations from our first-order analysis, the developed mathematical framework presents an important building block for further in-depth studies focused on a specific scenario or setup. Possible future work directions in this area may include: (i)~tailoring the antenna radiation model to a specific type of antenna in use (e.g., horn, lens, array, or reflecting surface); (ii)~accounting for inter-orbit cross-links; and (iii)~modeling available interference mitigation techniques and advanced deployments (e.g., by using channelization or coordination among the satellite transmissions), among other promising extensions.

In this context, this article also facilitates a new research avenue, particularly in developing intelligent routing and medium access protocols, due to the time-dependent nature of interference. Additionally, exploring the identified trade-offs in standardization, policy, and regulatory contexts could lead to solutions for interference-free coexistence in next-generation high-rate satellite communications.

\balance
\vspace{-1.5mm}
\section*{Acknowledgment}
\vspace{-1mm}
This work has been supported in part by the projects CNS-1955004, CNS-2332721, and CNS-2225590 by the U.S. National Science Foundation (NSF), the project FA9550-23-1-0254 by the U.S. Air Force Office of Scientific Research (AFOSR), and by a fellowship from ``la Caixa'' foundation (ID 100010434, LCF/BQ/AA20/11820041).


\appendices{}
\vspace{-1mm}
\section{Relative angular offset between orbits}
\label{app:rel_angular_offset}
The relative angular offset $\Delta\beta$ is defined as the relative angular position of the satellites in two different orbits, which in the case of circular orbits, corresponds to the difference in the mean anomaly of the closest pairs of satellites in different orbits. In a setup where the two orbits have the same altitude, the relative angular offset is constant through time, since both orbits rotate at the same speed. However, if the two orbits have different altitudes, the relative angular offset between them will change over time, $\Delta\beta(t)$, since both orbits will have different periods. Concretely, we can obtain $\Delta\beta$ as:
\vspace{-1mm}
\begin{equation}\label{eq:rel_angular_offset}
    \Delta\beta(t) = (\omega_1-\omega_2)t,
    \vspace{-1mm}
\end{equation}
where $\omega_1$ and $\omega_2$ are the angular speeds from the two different orbits, respectively, which for circular orbits are assumed to be constant. We can express the two orbital angular speeds as a function of the rotational period ($\omega=\frac{2\pi}{T}$) and relate the period to the orbital parameters with Kepler's Third law:
\vspace{-2mm}
\begin{equation}\label{eq:third_kepler_law}
    (R_{\textrm{E}}+h)^3 / T^2 = \mu / 4\pi^2,
    \vspace{-1mm}
\end{equation}
where $\mu\approx 3.986 \times 10^{14}\  m^2 s^{-2}$ is the Earth's standard gravitational parameter. For a two-orbit setup, the periodicity of the relative position between the satellites in each orbit and, therefore, the periodicity of the expected interference patterns, corresponds to a relative angular offset equal to the minimum angular distance between satellites, that is:
\vspace{-1mm}
\begin{equation}\label{eq:inc_beta_max}
    \Delta\beta_{\text{max}} = \min\left(2\pi / N_1, 2\pi / N_2\right),
    \vspace{-1mm}
\end{equation}
where, in this case, $N_1$ and $N_2$ correspond to the number of satellites in orbits 1 and 2, respectively. We can therefore obtain a closed expression for the period duration of the expected interference patterns by including \eqref{eq:inc_beta_max} and \eqref{eq:third_kepler_law} in \eqref{eq:rel_angular_offset}. Solving for $t$ we obtain:
\vspace{-1mm}
\begin{equation}
    T_{\Delta\beta_{\text{max}}} = \frac{\min\left(N_1^{-1}, N_2^{-1}\right)}{\sqrt{\mu} \left((R_{\textrm{E}}+h_1)^{-\frac{3}{2}} - (R_{\textrm{E}}+h_2)^{-\frac{3}{2}}\right)},
    \vspace{-1mm}
\end{equation}
where $h_1$ and $h_1$ are the orbital altitudes for each of the considered orbits, respectively.

\begin{samepage}
\bibliographystyle{IEEEtran}
\bibliography{final_version}

\vspace{-10mm}
\begin{IEEEbiography}[{\includegraphics[width=1in,height=1.25in,clip,keepaspectratio]{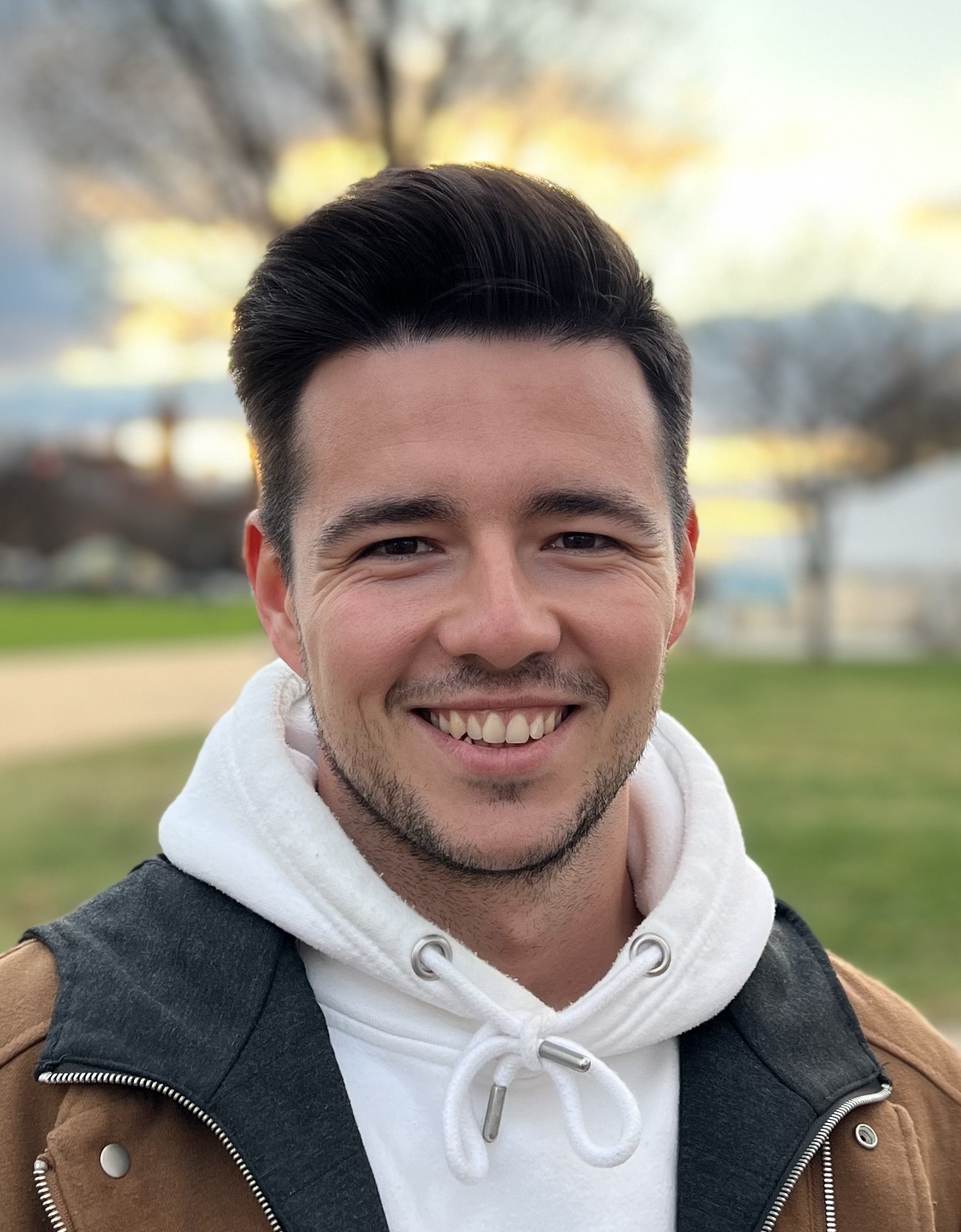}}]{Sergi Aliaga Torrens} is a PhD Candidate in the Ultra-broadband Nanonetworking Laboratory (UNLab) at the Institute for the Wireless Internet of Things (WIoT) at Northeastern University.  In 2021, under the CFIS program, he received two BSc degrees in Telecommunications Engineering and Industrial Engineering from the Polytechnic University of Catalonia (UPC), Spain. Earlier that year, he was a visiting student in the System Architecture Group at MIT. In 2020, he was awarded a La Caixa fellowship. During his graduate studies, Sergi has worked as a researcher at SES Satellite, received the best paper award in the Second IEEE Workshop on Non-Terrestrial-Networks for 6G, and participated in the 2023 Concurrent Engineering Workshop at the European Space Agency. His research interests include Satellite Communications in the THz band, Signal Processing for Joint Communication and Atmospheric Sensing, Propagation Modelling of THz waves in Space, and Constellation Design for the Next Generation of Non-Terrestrial Networks.
\end{IEEEbiography}

\begin{IEEEbiography}[{\includegraphics[width=1in,height=1.25in,clip,keepaspectratio]{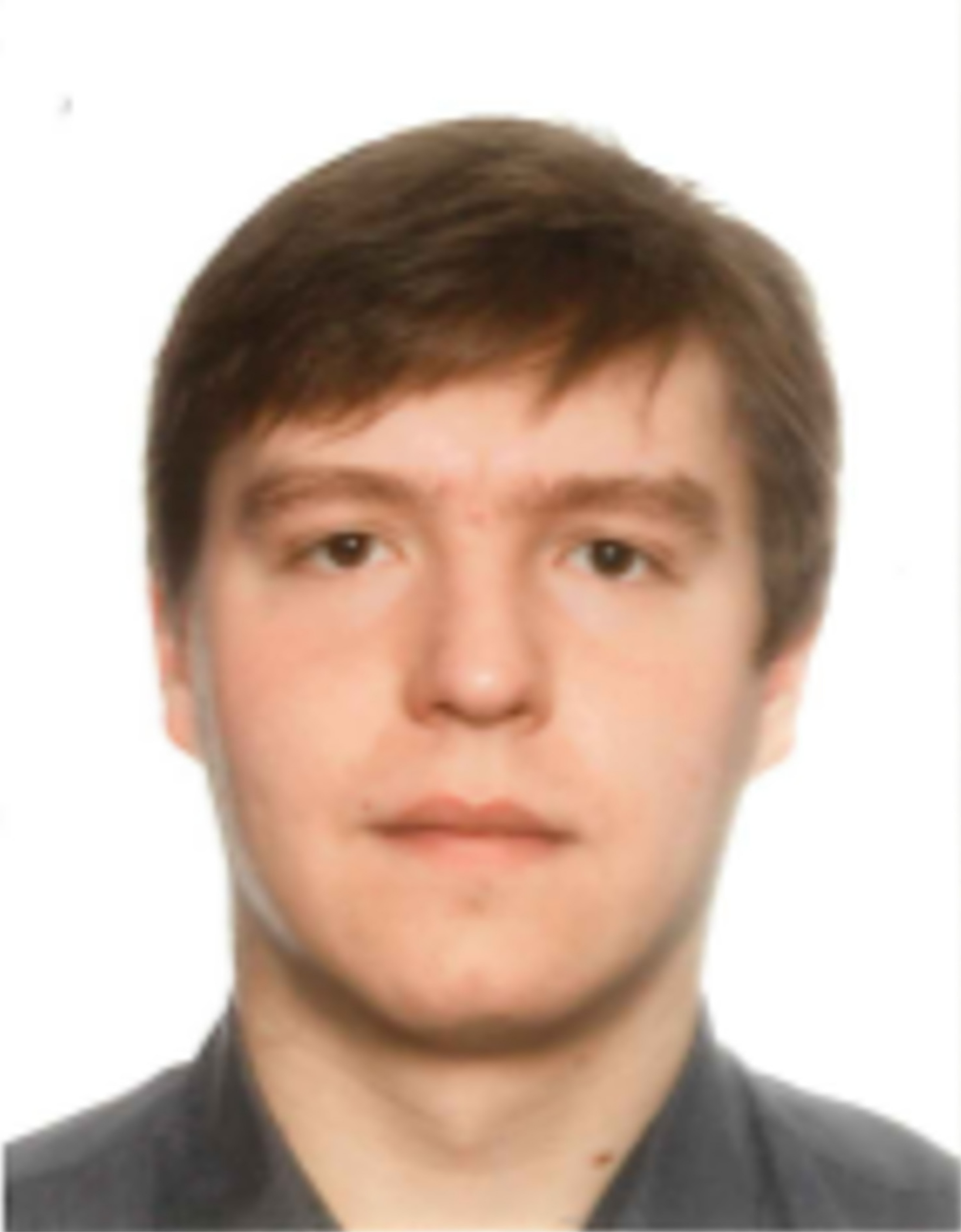}}] {Vitaly Petrov } is a Principal Research Scientist at Northeastern University, Boston, MA, USA. He obtained his M.Sc. degree in Information Systems Security from SUAI University, St.~Petersburg, Russia, in 2011, his M.Sc. degree in IT and Communications Engineering from Tampere University of Technology, Tampere, Finland, in 2014, and his Ph.D. degree in Communications Engineering from Tampere University, Finland, in 2020. Before joining Northeastern University in 2022, he was a Senior Standardization Specialist and a 3GPP RAN1 delegate with Nokia Bell Labs and later Nokia Standards. During his Ph.D. studies, he was a visiting researcher with the University of Texas at Austin, Georgia Institute of Technology, and King’s College London. His current research interests include terahertz band communications and networking. He is a recipient of the Best Student Paper Award at IEEE VTC-Fall 2015, the Best Student Poster Award at IEEE WCNC 2017, and the Best Student Journal Paper Award from IEEE Finland in 2019.
\end{IEEEbiography}

\begin{IEEEbiography}[{\includegraphics[width=1in,height=1.25in,clip,keepaspectratio]{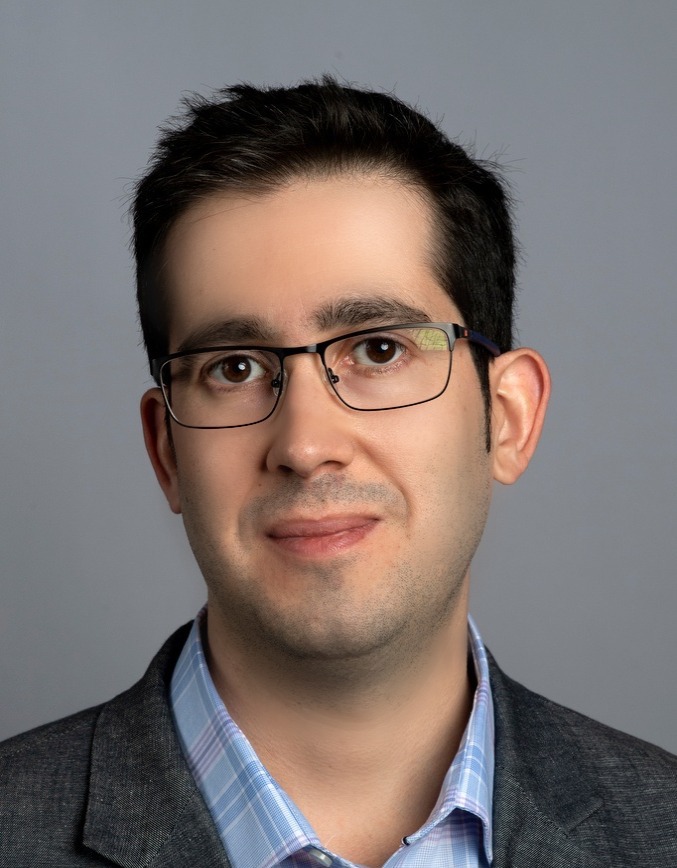}}]{Josep Miquel Jornet}
(M'13--SM'20--F'24) is a Professor in the Department of Electrical and Computer Engineering, the director of the Ultrabroadband Nanonetworking (UN) Laboratory, and the Associate Director of the Institute for the Wireless Internet of Things (WIoT) at Northeastern University (NU). He received a Degree in Telecommunication Engineering and a Master of Science in Information and Communication Technologies from the Universitat Politècnica de Catalunya, Spain, in 2008. He received his Ph.D. degree in Electrical and Computer Engineering from the Georgia Institute of Technology, Atlanta, GA, in August 2013. Between August 2013 and August 2019, he was in the Department of Electrical Engineering at the University at Buffalo (UB), The State University of New York (SUNY). He is a leading expert in terahertz communications, in addition to wireless nano-bio-communication networks and the Internet of Nano-Things. In these areas, he has co-authored more than 240 peer-reviewed scientific publications, including one book, and has been granted five US patents. His work has received over 16,000 citations (h-index of 58 as of December 2023). He is serving as the lead PI on multiple grants from U.S. federal agencies including the National Science Foundation, the Air Force Office of Scientific Research, and the Air Force Research Laboratory as well as industry. He is the recipient of multiple awards, including the 2017 IEEE ComSoc Young Professional Best Innovation Award, the 2017 ACM NanoCom Outstanding Milestone Award, the NSF CAREER Award in 2019, the 2022 IEEE ComSoc RCC Early Achievement Award, and the 2022 IEEE Wireless Communications Technical Committee Outstanding Young Researcher Award, among others, as well as six best paper and demo awards. He is a 
Fellow of the IEEE (Class of 2024) and an IEEE ComSoc Distinguished Lecturer (Class of 2022-2023). He is also the Editor in Chief of the Elsevier Nano Communication Networks journal and Editor for IEEE Transactions on Communications and Nature Scientific Reports.
\end{IEEEbiography}

\end{samepage}

\end{document}